\documentclass[10pt,twocolumn]{article}
\usepackage[margin=0.75in,columnsep=0.25in]{geometry}
\usepackage[T1]{fontenc}
\usepackage{lmodern}
\usepackage{amsmath,amssymb,bm}
\allowdisplaybreaks
\usepackage{graphicx}
\usepackage{booktabs}
\usepackage{multirow}
\usepackage{caption}
\usepackage{float}
\usepackage{dsfont}
\usepackage[dvipsnames]{xcolor}
\usepackage[numbers,sort&compress]{natbib}
\usepackage[colorlinks=true,linkcolor=blue!50!black,citecolor=blue!50!black,urlcolor=blue!50!black]{hyperref}
\title{Insect-inspired distributed reinforcement learning for active flow control}
\author{Kirubaharan Natarajan\thanks{Corresponding author: kirubaharan@vt.edu}~and Anne E. Staples\\[4pt]
\small Department of Mechanical Engineering, Virginia Tech, Blacksburg, VA 24061, USA}
\date{}

\begin{document}

\twocolumn[
\begin{@twocolumnfalse}
\maketitle
\begin{abstract}
Reinforcement learning (RL) has shown promise for flow control, but many existing approaches employ centralized neural-network policies, while modular and mixture-of-experts architectures do not necessarily yield interpretable functional organization or robust control. Here, we introduce an insect-inspired distributed RL architecture that combines functionally differentiated local controllers, memory, and learned sparse arbitration to coordinate control actions. Motivated by the distributed organization of insect neural circuits, the architecture is evaluated on two substantially different flow-control problems, with control intervals differing by nearly five orders of magnitude. The first benchmark is a three-zone thermally coupled recirculation network with a pump, a heater, and three branch valves that regulate the zone temperatures, with hidden disturbances. The second benchmark is a HydroGym's two-dimensional supersonic shock-vector-control nozzle under the inviscid, compressible flow assumption. In the thermal benchmark, over 10 seeds and 400 training updates, against parameter-matched MLP, GRU, and mixture of experts baselines, the Insect policy reduced the median mean zone temperature error to 1.5K (40-60\% below the baselines). In addition, reached the $\pm2$~K goal in 29\% of the episodes, compared to 1\% for baselines, while maintaining low recirculation risk. Removing its insect-inspired components together returned it to the baseline level. In the nozzle benchmark, over 5 seeds and 40,000 actions, against a parameter-matched MLP, the insect-inspired policy held the 5$^\circ$ with a steady-state error of 0.22$^\circ$ compared to Hydrogym's baseline MLP 0.94$^\circ$ while keeping the exit-plane pressure about eight times steadier. The findings suggest that functionally differentiated local controllers combined through learned arbitration offer a promising architectural approach to RL-based active flow control across disparate physical regimes and timescales.
\end{abstract}
\vspace{1.5em}
\end{@twocolumnfalse}
]
{\renewcommand{\thefootnote}{\fnsymbol{footnote}}\footnotetext[1]{Corresponding author: kirubaharan@vt.edu}}

\section{Introduction}

When Claude Shannon introduced his small robotic mouse, Theseus, in 1950, thinking machines were still mostly a subject for science fiction. Theseus itself was simply a magnet on wheels running through a maze with the electromagnetic relay switches beneath it that were Shannon's real construction \citep{shannon1951maze,soni2017mind}. Starting from a fresh and never-before-seen layout, Theseus explored the maze by trial and error while the relays recorded the paths that led closer to the end. On later runs, it used this stored information to go straight to the brass wedge of cheese waiting for it at the end. So, Theseus will have to try a few trials, learn its path through relays, and remember it to act accordingly later on; reaching its reward at the end of the maze is an early example of the trial-and-error learning on which today's reward-based methods are built. Historically, however, Theseus was not the first autonomous robot. In 1948, William Grey Walter built Elmer and Elsie, two biologically inspired ``tortoises'' that explored their surroundings using two sensors and two electronic nerve cells \citep{walter1950imitation}. Walter's work directly influenced generations of roboticists and became a cornerstone in the history of cybernetics, AI, and behavior-based robotics. However, it was Shannon's Theseus that brought the idea of a machine that learns from its own experience to a public audience \citep{time1952mouse}. Machine learning has since grown into a broad field that includes neural networks, reinforcement learning, and much of modern artificial intelligence.

Scientists have long been fascinated by the logic behind learning, understanding, remembering, processing, and classifying information inside brain So, these ways can be recreated in machines and used to perform demanding tasks. Reinforcement learning (RL) is one such development. It is inspired by how organisms learn from interactions with their environment to reach a goal. One important feature that makes RL different is reinforcement, or rewards. The organism, or the agent, either receives a reward for good actions or a penalty for bad ones and modifies its behavior accordingly to maximize the reward it accumulates over time \citep{sutton2018rl}. Deep reinforcement learning (DRL) extends this idea by replacing the lookup tables of classical RL with deep neural networks, which approximate the action values (or Q-values) or the policy directly from the inputs \citep{mnih2015dqn}. DRL is generally preferred when the system and its input data are rich, high-dimensional, and complex.

DRL has become a functional way to control closed-loop flows in fluid mechanics since \citet{rabault2019}; it was used to control synthetic jets in reducing the cylinder drag. The fast growth of this field has been extensively studied in the reviews by \citet{garnier2021review} and \citet{vignon2023review}, and platforms such as HydroGym \citep{lagemann2025hydrogym} now provide a compatible interface to 88 pre-configured CFD environments, which has further expedited the growth. HydroGym can be used elegantly to integrate the RL platform to understand active flow control. The standard environments provided across solvers vary from laminar wakes to compressible shock-vector control. In general, many such flow control studies use a single centralized network, either feedforward, like a multilayer perceptron (MLP), or with a recurrent layer like a gated recurrent unit (GRU), to cope with the partial observability \citep{ni2022pomdp}.

Flow control problems are often intricate combinations of several sub-tasks that require the controllers to provide suitable responses, like tracking a target point, avoiding a large transient, safely navigating through an unstable condition, bypassing an unsafe regime, and so on, while saving actuator efforts and energy in doing that. It will be computationally hard for a centralized policy to represent all these in one set of weights. Therefore, modular policies are used in such problems since they split the work across different specialists. Then the mixture of experts will learn how to weight each task. It is not yet settled whether modularity alone will be an ideal replacement for centralized policies in flow control problems. \citet{mittal2022modular} says that the routing must be very carefully designed, or else the network models often fail to specialize. 

Evidently, this unsettled debate can rest with a solution offered by nature! A recent exceptional study on the brain-and-nerve connectome of \emph{Drosophila} \citep{bates2026banc} shows that the fly's behavior is controlled by circuits distributed across the brain and nerve cord, rather than controlled by a single command center. Two brain structures are especially relevant here. The mushroom body stores the associations (for instance, if the fly smells lilies while being attacked or punished, it learns to avoid that odor later) using sparse Kenyon-cell codes \citep{lin2014sparse}, and the central complex keeps a persistent estimate of the fly's heading in a ring-attractor network \citep{seelig2015,kim2017ring}. For a long time, similar layered, arbitrated controllers have been built in the field of robotics \citep{brooks1986}.

Considering these elements, we have developed an RL policy, which we call the 'Insect' policy, and tested it on two flow control problems of very distinct characteristics. The first is a zero-dimensional, reduced-order thermally coupled recirculation network with three zones, each with its own branch experiencing hidden disturbances and hysteretic recirculation modes. The policy have to control the pump, the heater, and the three valves to manage the plant and run it under safe and efficient conditions where memory and switching behaviors should matter. The second one is a HydroGym's two-dimensional shock-vector-control nozzle \citep{lagemann2025hydrogym} where the policies control the secondary injectors downstream to deflect the thrust vector \citep{deere2003ftv}. This is an inviscid and compressible flow problem, with well-defined flow physics. The control interval of the first problem is 1~s, whereas for the second problem it's 15~$\mu$s. 

In summary, this work contributes an actor--critic architecture with a sparse, multi-timescale Kenyon cell memory, a ring-attractor state, a recurrent command center, four labeled local controllers, and a two-level arbiter. It is trained end-to-end with a recurrent PPO. The Insect architecture is compared against parameter-matched MLP, GRU, and mixture-of-experts baselines over 10 seeds in the thermal benchmark. For the nozzle benchmark, Insect policy is compared against HydroGym's reference PPO policy over 5 seeds.

\section{Insect policy architecture}
\label{sec:arch}

\begin{figure*}[t]
\centering
\includegraphics[width=0.8\textwidth]{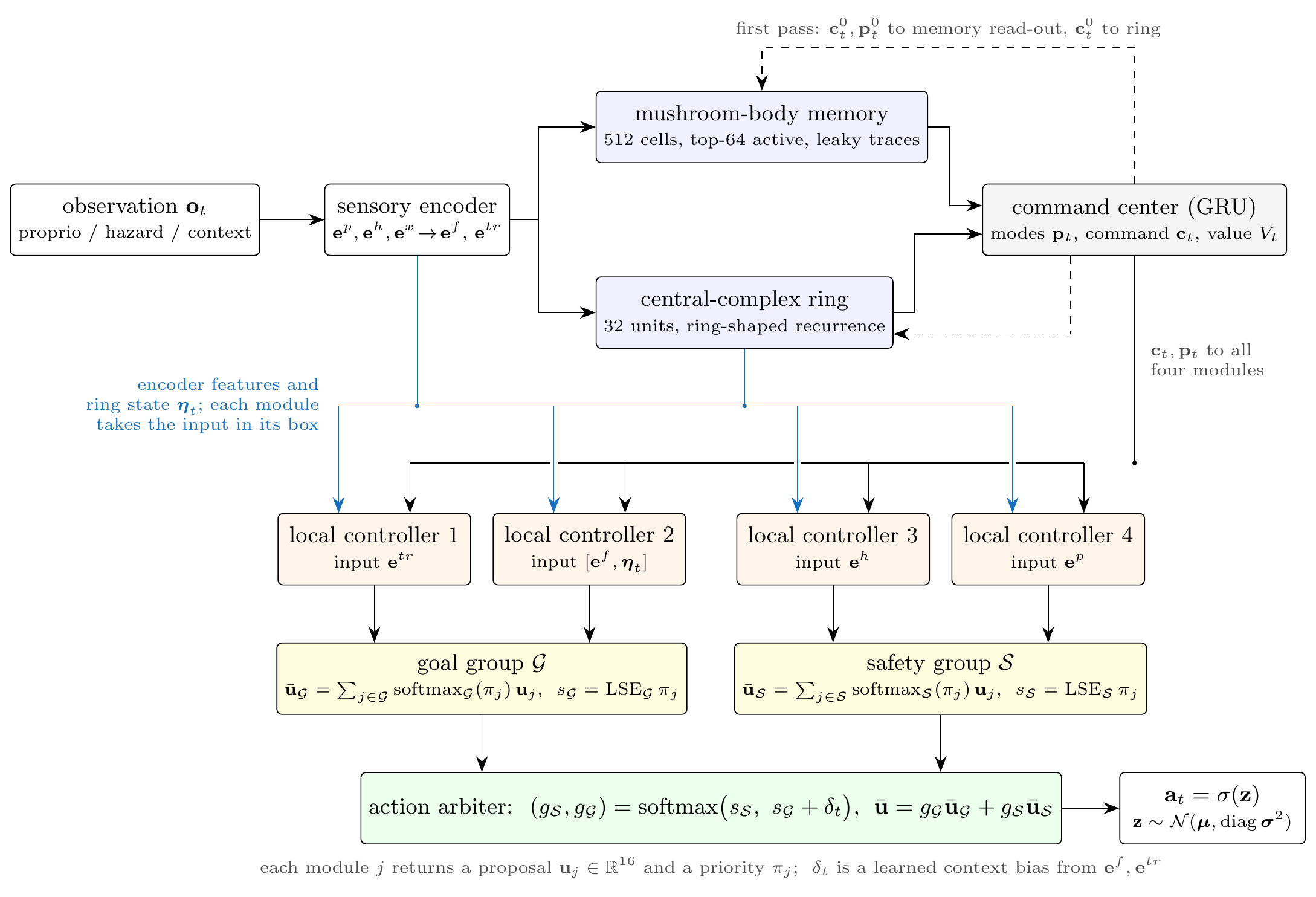}
\caption{Insect policy. Each local controller receives its own encoder features plus the command $\mathbf{c}_t$ and modes $\mathbf{p}_t$ from the command center and returns a proposal $\mathbf{u}_j$ and a priority $\pi_j$. The arbiter mixes the proposals within and then between the goal and safety groups. Only each controller's input is fixed; what it does is learned.}
\label{fig:arch}
\end{figure*}

Figure~\ref{fig:arch} summarizes the Insect policy. At the start of every control step, the policy receives an observation vector $\mathbf{o}_t$. This updates three recurrent states and outputs a bounded action vector $\mathbf{a}_t\in(0,1)^{n_a}$, which will be given to each actuator in the environment along with the value estimate $V_t$. Here ${n_a}$ is the number of actuators defined in the environment; for instance, the thermal network has 5, and the nozzle network has 2. 

\subsection{Sensory encoding}
The observation vector from the environment is split into a proprioceptive part $\mathbf{o}^p$ (for thermal benchmark: its temperatures, their increments, and the previous actions; for nozzle benchmark: its current and target thrust vector angles) and a hazard part $\mathbf{o}^h$ (for thermal benchmark: recirculation and pressure signals; for nozzle benchmark: the four downstream wall-pressure probe readings) and a context part $\mathbf{o}^x$ (for thermal benchmark: branch resistance deviations, in-band hold progress, and three zero-heat loads since they are not measured so the heat pulses stay hidden; for nozzle benchmark: none). For HydroGym's nozzle environment, we did not specify the context part, so the $\mathbf{e}^f=f_{\mathrm{fuse}}([\mathbf{e}^p;\mathbf{e}^h])$, and the tracking code is formed directly from the two observed angles (current \& target thrust angle) $\mathbf{e}^{tr}=\tanh(\mathbf{W}_{tr}\mathbf{o}^p+\mathbf{b}_{tr})$. Separate two-layer ReLU networks encode each part, and a further network fuses them:
\begin{align}
\mathbf{e}^p&=f_p(\mathbf{o}^p),\quad \mathbf{e}^h=f_h(\mathbf{o}^h),\quad \mathbf{e}^x=f_x(\mathbf{o}^x),\notag\\
\mathbf{e}^f&=f_{\mathrm{fuse}}\big([\mathbf{e}^p;\mathbf{e}^h;\mathbf{e}^x]\big),
\end{align}
with $\mathbf{e}^p,\mathbf{e}^x\in\mathbb{R}^{64}$ and $\mathbf{e}^h,\mathbf{e}^f\in\mathbb{R}^{128}$. A linear tracking code $\mathbf{e}^{tr}=\tanh(\mathbf{W}_{tr}[\tilde{\mathbf{e}};\Delta\mathbf{T}]+\mathbf{b}_{tr})\in\mathbb{R}^{128}$ is formed from the normalized zone errors $\tilde{\mathbf{e}}$ and their increments.

\subsection{Mushroom-body memory}
In Kenyon-like cells, the fused code from the sensory encoder is expanded into $K=512$ units. Here, the set of sensory inputs is mapped onto a much larger set of cells; hence the name "expansion layer". Of this $512$ units, only the $k=64$ most active are kept:
\begin{equation}
\bm{\kappa}_t=\mathrm{ReLU}(\mathbf{W}_\kappa\mathbf{e}^f_t),\qquad
\tilde{\bm{\kappa}}_t=\bm{\kappa}_t\odot\mathds{1}\big[\kappa_{t,j}\in\mathrm{top}_k(\bm{\kappa}_t)\big].
\end{equation}
Based on the chrono initialization \citep{tallec2018chrono}, each unit $K$ of Kenyon-like cells carries a leaky trace with its own learnable retention $\lambda_j=\sigma(\theta_j)$,
\begin{equation}
\mathbf{m}_t=\bm{\lambda}\odot\mathbf{m}_{t-1}+(\mathbf{1}-\bm{\lambda})\odot\tilde{\bm{\kappa}}_t,
\end{equation}
initialized such that the half-lives $\tau_j$ are log-spaced between 1 and 600 steps, $\lambda_j=2^{-1/\tau_j}$. The command center also provides a top-down input to the memory read-out (back to the mushroom body as in Figure~\ref{fig:arch}) so that whatever is read from the memory depends on the current command and mode:
\begin{equation}
\mathbf{y}_t=\tanh\big(\mathbf{W}_r\mathbf{m}_t+\mathbf{W}_v[\mathbf{c}^0_t;\mathbf{p}^0_t]+\mathbf{b}\big)\in\mathbb{R}^{64}.
\end{equation}

\subsection{Central-complex ring state}
A 32-unit state $\bm{\eta}_t$ is updated through a ring-shaped recurrent kernel $\mathbf{J}$,
\begin{equation}
\bm{\eta}_t=\mathrm{LN}\Big(\tanh\big(\mathbf{W}_\eta[\mathbf{e}^f_t;\mathbf{o}^p_t]+\mathbf{J}^{\top}\bm{\eta}_{t-1}+\mathbf{W}_c\,\mathbf{c}^0_t+\mathbf{b}_\eta\big)\Big)
\end{equation}
Here LN is layer normalization \citep{ba2016layernorm} and $\mathbf{c}^0_t$ is a command input similar to the velocity signal that shifts the heading bump in the fly. This heading bump is a cluster in the central ring that represents the fly's angular orientation, and as the fly changes its orientation, the cluster moves around the ring accordingly. Now each of the ring units in the cluster are connected through a trainable $32\times32$ weight matrix $\mathbf{J}$. Before training, each ring unit is given positive connections to its nearby neighbor (local-excitation) and negative connections to units farther away (global-inhibition). This pattern is found in the fly's heading circuit as \citep{kim2017ring}:
\begin{align}
J_{ij}&\propto\exp\!\left(-\frac{d_{ij}^2}{2(n_\eta/12)^2}\right)-0.15,\notag\\
d_{ij}&=\min\big(|i-j|,\,n_\eta-|i-j|\big).
\end{align}
The width $n_\eta/12$ and offset 0.15 are chosen specifically for this study. To avoid recurrent terms amplifying the ring state at initialization, the kernel is scaled to unit Frobenius norm, which bounds its spectral norm by one.
Neither of the benchmarks discussed in this work involves a heading, so $\bm{\eta}_t$ here simply serves as an extra recurrent state with ring design. We still kept the ring layout so the same architecture can be easily carried over to a new benchmark involving navigational problems. In that case, it would be essential to track heading. 

\subsection{Command center}
A GRU \citep{cho2014gru} with 128 units is stepped twice per control step. This resolves the age-long chicken-and-egg problem. The memory read-out and the ring update use the command center's command, $c$. At the same time, the command center computes $c$ from the memory read-out and the ring state. There comes a situation where they both need each other. So, we run the GRU two times every control step. 
At the first pass (depicted in Figure~\ref{fig:arch}), GRU uses the previous ring state and the memory read-out without the top-down (or valence) term since no commands exist yet:
\begin{align}
\mathbf{h}^0_t&=\mathrm{GRU}\big([\mathbf{e}^f_t;\bm{\eta}_{t-1};\mathbf{y}^0_t],\,\mathbf{h}_{t-1}\big),\notag\\
\mathbf{p}^0_t&=\mathrm{softmax}\big(\mathbf{W}_p\mathbf{h}^0_t/\tau_p\big),\qquad
\mathbf{c}^0_t=\tanh(\mathbf{W}_c\mathbf{h}^0_t).
\end{align}
These are used to update $\bm{\eta}_t$ and to form $\mathbf{y}_t$ with valence input $[\mathbf{c}^0_t;\mathbf{p}^0_t]$. The second pass will start from the state of the first pass $\mathbf{h}^0_t$ and then use the updated ring state and memory read-out to give the outputs used by the rest of the network,
\begin{align}
\mathbf{h}_t&=\mathrm{GRU}\big([\mathbf{e}^f_t;\bm{\eta}_t;\mathbf{y}_t],\,\mathbf{h}^0_t\big),\notag\\
\mathbf{p}_t&=\mathrm{softmax}(\mathbf{W}_p\mathbf{h}_t/\tau_p),\quad
\mathbf{c}_t=\tanh(\mathbf{W}_c\mathbf{h}_t),\notag\\
V_t&=\mathbf{w}_V^{\top}\mathbf{h}_t+b_V,
\end{align}
with $n_m=6$ modes, a 16-dimensional command and mode temperature $\tau_p=0.65$. The critic shares the same recurrent trunk.

\subsection{Local controllers and arbitration}
We added four local controllers $j\in\{\mathrm{tr},\mathrm{rt},\mathrm{sr},\mathrm{cs}\}$ that receive different inputs $\mathbf{x}_j$ along with the command center's command and mode probabilities $[\mathbf{c}_t;\mathbf{p}_t]$. In the thermal network, the local controllers are named for their intended roles: transport ($\mathbf{x}_{\mathrm{tr}}=\mathbf{e}^{tr}$), recover temperature ($\mathbf{x}_{\mathrm{rt}}=[\mathbf{e}^f;\bm{\eta}]$), suppress recirculation ($\mathbf{x}_{\mathrm{sr}}=\mathbf{e}^h$) and conserve ($\mathbf{x}_{\mathrm{cs}}=\mathbf{e}^p$). However, we did not provide handwritten rules on how to react to each situation. That will be totally from the learning. Each controller returns an action code and a priority,
\begin{align}
\mathbf{u}_j&=\tanh\big(g_j([\mathbf{x}_j;\mathbf{c}_t;\mathbf{p}_t])\big)\in\mathbb{R}^{16},\notag\\
\pi_j&=q_j([\mathbf{x}_j;\mathbf{c}_t;\mathbf{p}_t])\in\mathbb{R}.
\end{align}
The four local controllers are grouped further into two groups: goal and safety. The arbiter first weighs a goal group $\mathcal{G}=\{\mathrm{tr},\mathrm{rt}\}$ against a safety group $\mathcal{S}=\{\mathrm{sr},\mathrm{cs}\}$, then the members within each group:
\begin{align}
(g_{\mathcal{S}},g_{\mathcal{G}})&=\mathrm{softmax}\Big(\log\!\sum_{j\in\mathcal{S}}e^{\pi_j},\ \log\!\sum_{j\in\mathcal{G}}e^{\pi_j}+\delta_t\Big),\\
\delta_t&=\mathbf{w}_\delta^{\top}[\mathbf{e}^f_t;\mathbf{e}^{tr}_t]+b_\delta,\\
w_j&=g_{\mathcal{K}(j)}\,\frac{e^{\pi_j}}{\sum_{k\in\mathcal{K}(j)}e^{\pi_k}},\qquad
\bar{\mathbf{u}}_t=\sum_j w_j\mathbf{u}_j,\\
\bm{\mu}_t&=\mathbf{W}_\mu\tanh(\mathbf{W}_a\bar{\mathbf{u}}_t+\mathbf{b}_a)+\mathbf{b}_\mu,
\end{align}
where $\mathcal{K}(j)$ is the group containing local controller $j$. The context bias $\delta_t$ is initialized to zero, such that the two-level rule reduces to a single softmax over the four priorities. The training can then learn to shift weight between the groups on the encoded state. The weights $w_j$ are shared by all actuators. The vector $\bm{\mu}_t$ is the mean action, one entry per actuator, before it is squashed into $(0,1)$ by the action distribution.

\subsection{Action distribution}
To keep every command inside $(0,1)$, the actions are drawn from a logistic-normal distribution:
\begin{equation}
\mathbf{z}\sim\mathcal{N}\big(\bm{\mu}_t,\mathrm{diag}(\bm{\sigma}^2)\big),\qquad \mathbf{a}_t=\sigma(\mathbf{z}),
\end{equation}
\begin{align}
\log\pi(\mathbf{a}_t)=\sum_i\Big[&\log\mathcal{N}\big(\mathrm{logit}\,a_{t,i};\mu_{t,i},\sigma_i^2\big)\notag\\
&-\log a_{t,i}-\log(1-a_{t,i})\Big].
\end{align}
The standard deviation $\bm{\sigma}$ is a state-independent parameter clipped to $[0.08,\,1]$ and initialized at 0.5 (again set specifically for the two benchmarks discussed here). The output bias $\mathbf{b}_\mu$ is set to $\mathrm{logit}(\mathbf{a}_0)$, with $\mathbf{a}_0=(0.45,0.85,0.5,0.5,0.5)$ for (pump, heater, and 3 valves), and $\mathbf{W}_\mu$ is initialized orthogonally with gain 0.01. Deterministic evaluation uses $\mathbf{a}_t=\sigma(\bm{\mu}_t)$.

\subsection{Baselines}
\label{sec:baselines}
All thermal baselines use the same logistic-normal head, output-bias initialization and standard-deviation bounds to keep a fair ground. Their hidden widths were chosen so that the trainable parameter counts match the Insect policy to within 0.1\% (Table~\ref{tab:params}).
\begin{itemize}
\item \textbf{Centralized MLP}: three hidden ReLU layers of width 481; policy mean and value read from the same trunk; no memory.
\item \textbf{Centralized GRU}: a two-layer ReLU encoder and a GRU cell, both of width 259. With one linear layer on GRU state giving the action, while the other gives the value. It utilizes its memory capability to control.
\item \textbf{Mixture of experts (MoE)}: four MLP experts (two hidden layers of width 302), each proposing an action mean, mixed by a softmax router (one hidden layer of 64) that sees the full observation. Has a separate critic (256, 256). No memory and no role labels. The router receives the same entropy term as the Insect arbiter (Section~\ref{sec:ppo}). It provides learned modular routing.
\end{itemize}

\begin{table*}[t]
\centering
\caption{Trainable parameters. Thermal counts are from the training log of job 656091; nozzle counts from the matched-width configuration used for the step-target runs.}
\label{tab:params}
\begin{tabular}{llr}
\toprule
Problem & Policy & Parameters \\
\midrule
\multirow{4}{*}{Thermal network} & Insect & 479\,686 \\
 & Centralized MLP (3 $\times$ 481) & 479\,568 \\
 & Centralized GRU (259) & 479\,938 \\
 & MoE (4 experts, 2 $\times$ 302) & 479\,654 \\
\midrule
\multirow{2}{*}{HydroGym nozzle} & Insect & 464\,544 \\
 & SB3 PPO MLP (actor and critic 2 $\times$ 477) & 464\,126 \\
\bottomrule
\end{tabular}
\end{table*}

\section{Benchmark 1: thermally coupled recirculation network}
\label{sec:thermal}

\begin{figure}[!htbp]
\centering
\includegraphics[width=\linewidth]{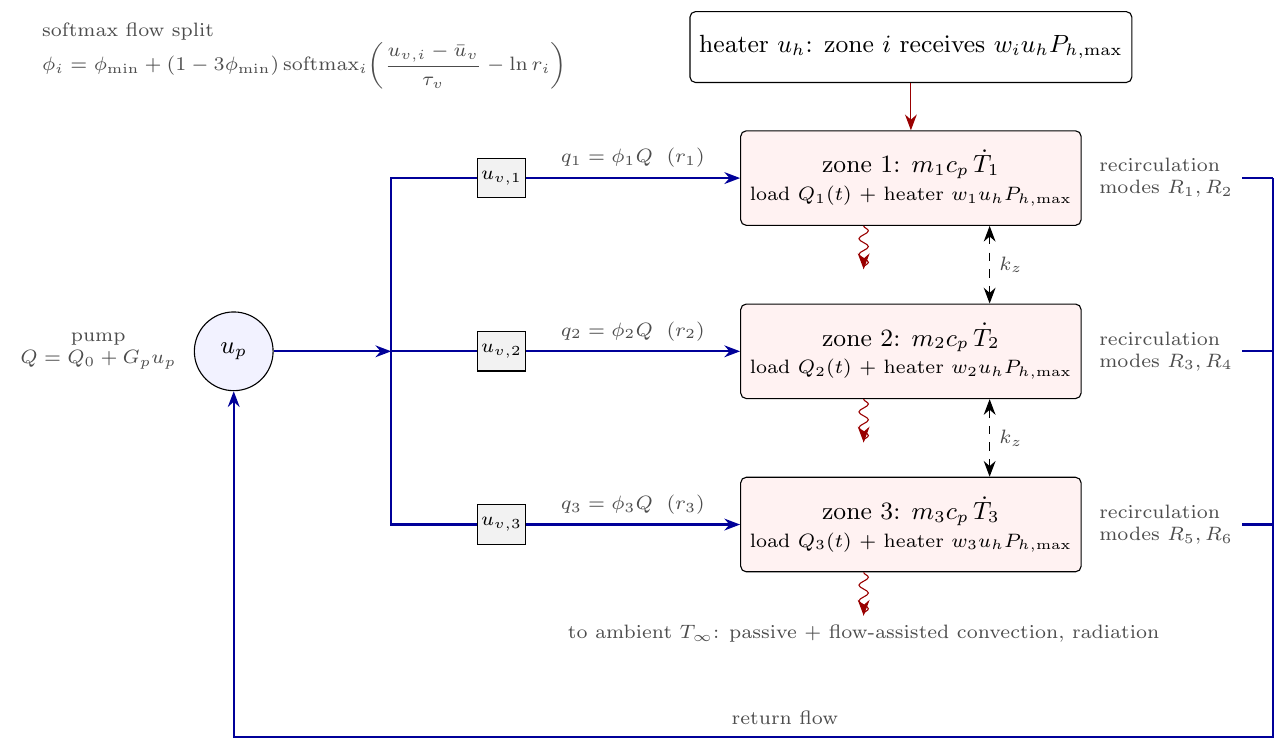}
\caption{Thermal recirculation network. A pump drives flow $Q$ that a softmax valve law splits into three branches, each cooling one thermal zone. Each branch carries two hysteretic recirculation modes that reduce its heat-transfer effectiveness. The heater distributes power to all zones with fixed weights; zones exchange heat with each other and with the ambient.}
\label{fig:thermal}
\end{figure}

The benchmark used here is a thermal management plant with three thermal zones cooled by a pumped water loop. This is a skeletal model with no real physics that was entirely designed by ourselves for this study. This lumped-parameter model is a reduced-order control benchmark, not a CFD model. As mentioned before, all the flow quantities are in normalized units. The policy can control the pump $u_p$, the heater $u_h$, and three branch valves $u_{v,i}$ to optimally run the plant and remove the heat from the three zone effectively. All the control commands from the policies are compressed to be between 0 and 1 and are allowed to vary them smoothly. Key parameters are listed in Table~\ref{tab:thermalparams}.

\subsection{Energy balance model}
Each zone in the benchmark,  $i\in\{1,2,3\}$ has mass $m_i$ and temperature $T_i$. The lumped-parameter heat balance for each zone is given by:
\begin{equation}
\begin{split}
\underbrace{m_i c_p\frac{dT_i}{dt}}_{\substack{\text{rate of change}\\\text{of stored heat}}}
={}&\underbrace{Q_i(t)}_{\substack{\text{external}\\\text{heat load}}}
+\underbrace{w_i u_h P_{h,\max}}_{\substack{\text{heater}\\\text{power}}}
-\underbrace{h_p\,\Theta_i}_{\substack{\text{passive}\\\text{loss}}}
-\underbrace{h_{f,i}\,\Theta_i}_{\substack{\text{cooling by}\\\text{coolant flow}}}\\[4pt]
&-\underbrace{\epsilon\sigma_{SB}\frac{A}{3}\big(T_i^4-T_\infty^4\big)^{+}}_{\text{radiation to the surroundings}}
+\underbrace{k_z\big(\bar T-T_i\big)}_{\substack{\text{heat exchange}\\\text{between zones}}},
\end{split}
\label{eq:energy}
\end{equation}
where $\Theta_i=\max(T_i-T_\infty,0)$, $\bar T$ is the mean zone temperature, and $(\cdot)^+=\max(\cdot,0)$. The passive and flow-assisted conductances are
\begin{align}
h_p&=f_p\,c_{\mathrm{cool}}\,q_{\mathrm{ref}},\qquad
h_{f,i}=\frac{f_f\,c_{\mathrm{cool}}\,q_{\mathrm{ref}}\,(q_i/q_{\mathrm{ref}})^{0.8}}{1+\beta_R\,\rho_i^2},\notag\\
\rho_i&=\max_{k\in\mathcal{B}(i)}\frac{R_k}{R_{\mathrm{unsafe}}},
\end{align}
so flow through branch $i$ cools zone $i$ with a power-law dependence on the branch flow $q_i$, and recirculation in that branch ($\rho_i$) degrades the cooling. Here $\mathcal{B}(i)$ is the pair of recirculation modes attached to branch $i$ and $q_{\mathrm{ref}}=(Q_0+0.5G_p)/3$.

\subsection{Hydraulics}
The pump sets the total flow $Q=Q_0+G_p u_p$. Then the split is set by the valves and branch resistances $r_i(t)$ through a softmax:

\begin{align}
\phi_i&=\phi_{\min}+(1-3\phi_{\min})\,\frac{\exp\!\big((u_{v,i}-\bar u_v)/\tau_v-\ln r_i\big)}{\sum_{l}\exp\!\big((u_{v,l}-\bar u_v)/\tau_v-\ln r_l\big)},\notag\\
q_i&=\phi_i Q,
\label{eq:split}
\end{align}
where $\phi_{\min}$ is leakage, so only the relative valve openings matter and a degraded (high-resistance) branch draws less flow at equal openings. The pressure drop is
\begin{equation}
\Delta p=k_{dp}\sum_i r_i q_i^2+\sum_k k_{R,k}R_k.
\end{equation}

\begin{table*}[t]
\centering
\caption{Thermal-network parameters used in the final benchmark.}
\label{tab:thermalparams}
\small
\begin{tabular}{p{5.6cm}p{2.9cm}p{6.6cm}}
\toprule
Quantity & Symbol & Value \\
\midrule
Zone mass, specific heat & $m_i$, $c_p$ & 2/3~kg each, 4180~J\,kg$^{-1}$K$^{-1}$ \\
Ambient, target temperature & $T_\infty$, $T_{\mathrm{ref}}$ & 300~K, 342~K \\
Base heat loads & $Q_i^0$ & 125, 115, 120~W \\
Heat pulse & $\Delta Q$ & 1400~W $\times\,\mathcal{U}(0.95,1.10)$, six windows, rotating zone \\
Heater & $P_{h,\max}$, $w_i$ & 1200~W; 0.38, 0.32, 0.30 \\
Inter-zone conductance & $k_z$ & 18~W\,K$^{-1}$ \\
Cooling coefficient & $c_{\mathrm{cool}}$ & 120 $\times\,\mathcal{U}(0.92,1.05)$ \\
Passive and flow fractions; recirculation factor & $f_p$, $f_f$, $\beta_R$ & 0.133, 0.80, 1.50 \\
Radiation & $\epsilon$, $A$ & 0.8, 0.3~m$^2$ \\
Pump & $Q_0$, $G_p$ & 0.12, 1.05 (normalized flow) \\
Valve law & $\tau_v$, $\phi_{\min}$ & 0.2, 0.02 \\
Branch resistance & $\hat r_i$ & 1.00, 1.08, 0.96; one branch $\times\,\mathcal{U}(0.95,1.12)$ \\
Branch degradation & & $r_i\times1.85$ in five windows, rotating branch \\
Critical flow & $\hat Q_{c,k}$ & 0.66, 0.75, 0.69, 0.78, 0.66, 0.75 \\
Unsafe / crisis risk & $R_{\mathrm{unsafe}}$, $R_{\mathrm{crisis}}$ & 1.0, 1.55 \\
Recirculation gains & $\kappa_r,\kappa_{vu},\kappa_{pu}$, $\kappa_{ps},\kappa_{vs}$ & 2.80, 1.85, 2.10, 1.90, 1.60 \\
Hidden vulnerability time constant & $\tau_\eta$ & 240~s \\
Control step, episode limit & $\Delta t$, $T_{\mathrm{ep}}$ & 1~s, 900~s \\
\bottomrule
\end{tabular}
\end{table*}

\subsection{Recirculation dynamics}
In this benchmark, 'recirculation' is not a backflow but a hazard that grows when a branch is overloaded. There are six recirculation modes $R_k$, two per branch, that respond to flow in their branch. A mode becomes active ($\chi_k=1$) when the flow $q_{b(k)}$ exceeds a critical flow $Q_{c,k}(t)$. To provide hysteresis, the mode becomes inactive only when it falls below $Q_{c,k}-\Delta q_k$. 
\begin{equation}
\begin{split}
\frac{dR_k}{dt}={}&\underbrace{\alpha^{\mathrm{on}}_k\big(0.22\chi_k+1.05\,o_k^{1.25}\big)\big(1+0.55s^v_k+0.45s^p\big)}_{\text{growth}}\\[2pt]
&+\underbrace{\gamma_k\chi_k}_{\text{persistence}}
+\underbrace{c_kR_{k-1}}_{\substack{\text{coupling from}\\\text{the previous mode}}}
+\underbrace{\xi_k}_{\text{noise}}
-\underbrace{d_kR_k}_{\text{decay}},
\end{split}
\label{eq:recirc}
\end{equation}
\begin{align}
\xi_k&\sim\mathcal{N}(0,\sigma_\xi^2),\notag\\
d_k&=\alpha^{\mathrm{off}}_k\,\frac{0.30+\kappa_r\varrho_k+\kappa_{vu}\ell^v_k+\kappa_{pu}\ell^p}{1+\kappa_{ps}s^p+\kappa_{vs}s^v_k}
\end{align}
where the six dimensionless recirculation drivers are
\begin{alignat}{2}
&\text{overload:}\quad && o_k=\frac{(q_{b(k)}-Q_{c,k})^+}{Q_{c,k}},\notag\\[3pt]
&\text{relief:}\quad && \varrho_k=\frac{(Q_{c,k}-q_{b(k)})^+}{Q_{c,k}},\notag\\[3pt]
&\text{branch unloading:}\quad && \ell^v_k=\frac{(1/3-\phi_{b(k)})^+}{1/3},\notag\\[3pt]
&\text{branch sustaining:}\quad && s^v_k=\frac{(\phi_{b(k)}-1/3)^+}{1/3},\notag\\[3pt]
&\text{pump unloading:}\quad && \ell^p=\frac{(0.45-u_p)^+}{0.45},\notag\\[3pt]
&\text{pump sustaining:}\quad && s^p=\frac{(u_p-0.45)^+}{0.55}.
\label{eq:drivers}
\end{alignat}
In the neutral state, all these drivers are zero. Thus, the branch carries exactly its critical flow, the valves split the flow evenly ($\phi=1/3$), and the pump command is 0.45. Recirculation therefore clears faster when the policy unloads the affected branch and reduces the pump pressure. The persistence term in the recirculation model maintains the hazard if a policy forces flow through the vulnerable branch. $R_k$ is clipped to $[0,R_{\max,k}]$ with the unsafe level being $R_{\mathrm{unsafe}}=1$.

During these vulnerability windows, the critical threshold for the recirculation hazard drops, so it starts at even low flow:
\begin{equation}
Q_{c,k}(t)=\big(\hat Q_{c,k}-\delta_k\,\mathds{1}_{\mathrm{vul}}(t)\big)\,\nu_k(t).
\end{equation}
This is hidden from the policies. So the hidden multiplier $\nu_k\le 1$ is responsible for making one randomly chosen branch per episode vulnerable. Inside a window the hidden multiplier falls slowly (time constant $\tau_\nu=240$~s) toward a random level $m_b\sim\mathcal{U}(0.60,1.0)$, and on the outside it recovers toward 1. Neither $\nu_k$ nor $m_b$ is observed, making this a hidden disturbance to the plant.

Equations~\eqref{eq:energy}--\eqref{eq:recirc} are integrated with the explicit Euler method in two substeps of 0.5~s, where we hold the action, flows, heat loads, and critical flows fixed and update the temperatures and recirculation risks.

\subsection{Disturbances and observability}
At the beginning of every episode, there will be a deliberate upset in the temperature of the thermal zones: the three zones begin at 4--9~K away from the target, with at least one zone too hot and one too cold, so there is an 8~K spread of temperature variation from the target. The policies have to use their capabilities to bring the temperature closer to the target (Figure~\ref{fig:recovery}). During the 900~s episode, the plant experiences six heat pulses and vulnerability windows and five branch-degradation windows at random time intervals. The magnitude of the heat pulse, cooling coefficient, resistance of one branch, and the vulnerability multiplier per episode are given in Table~\ref{tab:thermalparams}. With the sensor delay of 0 or 1 step, only 5\% of the episodes are trained under extreme ends of these ranges. To provide a fair chance and to train all four architectures on the same scenarios, the disturbance sequence is pre-generated for a given training seed.

The 26-D observations that the policies can see are the normalized zone errors $(T_i-T_{\mathrm{ref}})/5$~K, their increments, the previous action, the normalized recirculation risks $R_k/R_{\max,k}$, the mean risk increment, the normalized pressure drop, the branch-resistance deviations, $r_i/\hat r_i-1$, and the progress of the current in-band hold. The zone heat loads are not measured (their slots are zero), and the vulnerability state, cooling coefficient, and sensor delay are hidden. Temperature, risk, and pressure readings pass through the sensor delay.

\subsection{Reward and episode endings}
The per-step reward, scaled by 0.1, is
\begin{align}
r_t={}&-\Big[\underbrace{w_T\tfrac13\sum_i\big(\tfrac{e_i}{5}\big)^2}_{\text{mean error}}
+\underbrace{w_w\big(\tfrac{\max_i|e_i|}{5}\big)^2}_{\text{worst zone}} \nonumber\\[2pt]
&+\underbrace{w_u\big(\tfrac{\sigma_T}{2}\big)^2}_{\text{uniformity}}\Big]
-\underbrace{w_o\sum_i\big(\tfrac{e_i^+}{5}\big)^2}_{\text{overheating}} \nonumber\\[2pt]
&-\underbrace{w_R\Big[\rho_s^2+\tfrac14\overline{\rho_{s,k}^2}\Big]\nu_t}_{\text{recirculation risk}}
-\underbrace{w_{dp}\frac{\Delta p}{\Delta p_{\mathrm{ref}}}}_{\text{pressure drop}} \nonumber\\[2pt]
&-\underbrace{w_{pe}u_p^2}_{\substack{\text{pump}\\\text{energy}}}
-\underbrace{w_{he}u_h^2}_{\substack{\text{heater}\\\text{energy}}}
-\underbrace{w_{ve}\overline{(\Delta\phi_i)^2}}_{\substack{\text{flow-split}\\\text{changes}}} \nonumber\\[2pt]
&-\underbrace{w_s\Big[(\Delta u_p)^2+(\Delta u_h)^2+\tfrac12\overline{(\Delta u_{v,i})^2}\Big]}_{\text{command smoothness}} \nonumber\\[2pt]
&-\underbrace{c_{pr}\sum_k n_k}_{\substack{\text{staying}\\\text{unsafe}}}
-\underbrace{c_{cr}\sum_k\mathds{1}[R_k>R_{\mathrm{crisis}}]}_{\text{crisis}} \nonumber\\[2pt]
&+\underbrace{w_{\Phi}\big(\gamma\Phi_{t+1}-\Phi_t\big)}_{\substack{\text{shaping: reward for}\\\text{reducing the error}}}
+\underbrace{0.1\,\mathds{1}_{\mathrm{band}}}_{\substack{\text{in-band}\\\text{bonus}}},
\label{eq:reward}
\end{align}
where $e_i=T_i-T_{\mathrm{ref}}$ in kelvin, $\sigma_T$ is the spread of the zone temperatures, $\rho_s=(\max_k R_k/R_{\mathrm{unsafe}}-0.75)^+/0.25$ is a soft risk measure (with $\rho_{s,k}$ its per-mode version), $\nu_t=1.2$ inside vulnerability windows and 1 otherwise, $n_k$ counts consecutive unsafe steps of mode $k$, and $\mathds{1}_{\mathrm{band}}$ indicates that all zones are within $\pm2$~K with $\max_k R_k<0.8$. The term $w_\Phi(\gamma\Phi_{t+1}-\Phi_t)$ is potential-based shaping \citep{ng1999shaping} with $\Phi$ equal to the negative of the first bracket. The weights are $w_T=0.70$, $w_w=1.00$, $w_u=0.25$, $w_o=0.75$, $w_R=2.5$, $w_{dp}=0.075$, $w_{pe}=0.009$, $w_{he}=w_{ve}=0.010$, $w_s=0.008$, $c_{pr}=0.03$, $c_{cr}=3.5$, $w_\Phi=0.25$ and $\gamma=0.99$.

An episode \emph{succeeds} when all zones stay within $\pm2$~K with $\max_k R_k<0.8$ for 60~s. And at least 30~s of which overlaps a heat pulse, and no trip condition is active; a terminal bonus of 150 is then added. Otherwise it ends at 900~s. Trip conditions ($|e_i|>12$~K or $R_k>1.55$) are recorded but do not end the episode; the reward already penalizes large temperature errors and crisis-level recirculation.

\begin{figure*}[t]
\centering
\includegraphics[width=0.9\textwidth]{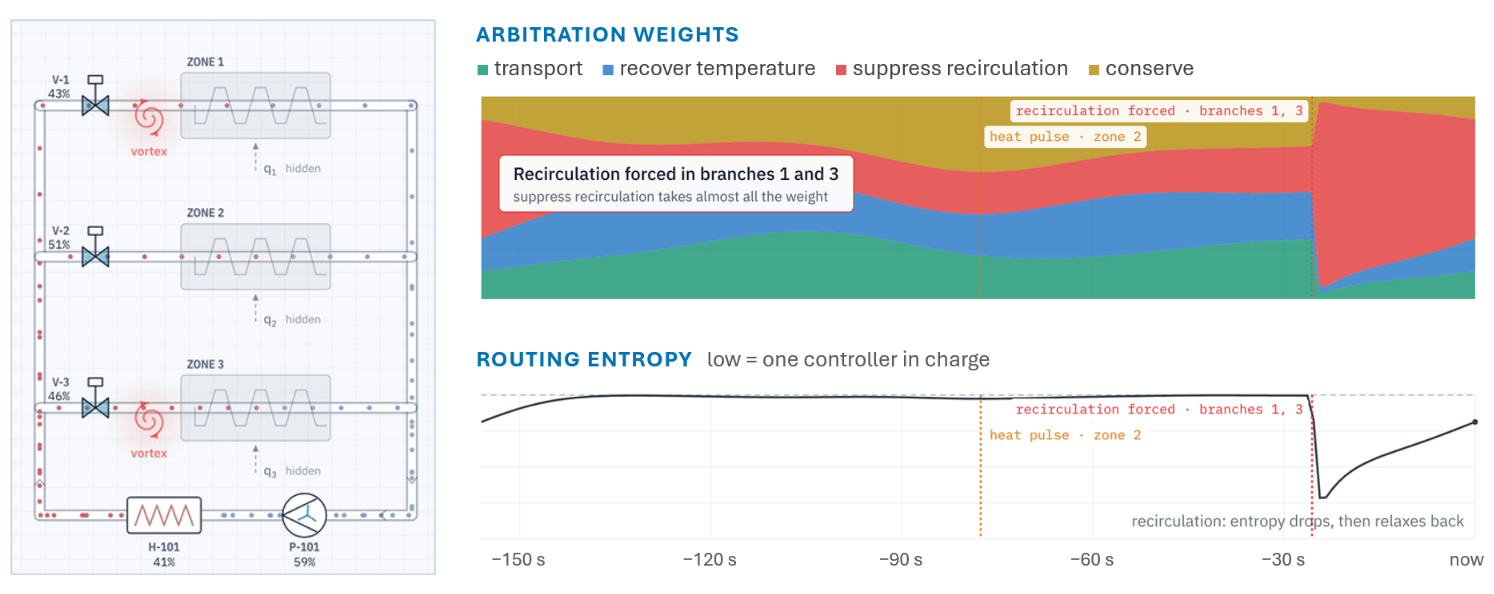}
\caption{Schematic of how the arbiter shares control among the four local controllers in the thermal benchmark (an illustration, not measured data). Left: the three-zone loop with valves V-1 to V-3, heater H-101 and pump P-101; the heat loads $q_1$--$q_3$ are not observed. The vortex symbols mark branches at risk of recirculation. Right: module weights and their entropy over time. A forced recirculation in branches 1 and 3 moves almost all of the weight to the suppress-recirculation controller; the entropy drops and then relaxes as the risk decays. The measured behavior of a trained policy is shown in Figure~\ref{fig:modules}.}
\label{fig:arb_schematic}
\end{figure*}

\subsection{Local Controllers}

Each local controller proposes an action at every step by looking at its own input; for instance, local controller 1 receives a prescribed input that helps it obtain action commands that meet that controller's purpose. Then the local controller outputs two things: a proposal and a priority. The proposal contains $16$ for each controller in this benchmark that are not actuator commands yet but are the controller's suggestion in an internal code. The priority helps the arbiter decide how much control is needed right now; it contains just $1$ number. As discussed above, the arbiter turns the four priorities into weights, blends the proposals accordingly, and decodes the blend into five actuator commands. The simple schematic of how this is handled based on the situation or hazard in a plant is shown in Figure~\ref{fig:arb_schematic}. So the goal is to make the insect-inspired policy work the same way as it did with the navigational benchmark (prior work with this same neural network) \citep{staples2026insect}. Thus giving each local controller of the insect-inspired network its own `sense.'

\subsubsection{Transport}

This controller observes how far each zone temperature is from the target temperature, along with the rate of change in the temperature. It gets the 128-D tracking code, which is built from the $3$ zone temperature errors and their respective rates of change. The purpose of this controller is to bring the zone temperature closer to the target. Therefore, it falls within the goal group of the arbiter.

\subsubsection{Suppress Recirculation}

This local controller gets the 128-D hazard code, which is built from the $8$ hazard readings, including the risk of each of the $6$ recirculation modes in three branches, the total risk rate, and the pressure drop. The purpose of this controller is to keep the branches away from recirculation. Therefore, it falls within the safety group of the arbiter. 

\subsubsection{Recover Temperature}

This gets a 128-D fused code along with the 32-D ring state, $160$ in total. This input is built from all the 26-D observation vector, including temperatures, hazards, and context, along with the central-complex ring state. The purpose is to coordinate a stronger recovery after large errors and instabilities. This falls within the goal group of the arbiter.

\subsubsection{Conserve}

Finally, this controller gets a 64-D proprioceptive code, which is built from $3$ temperature errors, their $3$ rates, and the previous commands (pump, heater, and $3$ valves). The purpose of this controller is to hold quiet, steady, and low-energy operation when the plant is safe. Therefore, it falls within the safety group of the arbiter. 

\section{Benchmark 2: HydroGym shock-vector-control nozzle}
\label{sec:nozzle}

\subsection{Working and Governing equations}
\begin{figure}[!htbp]
\centering
\includegraphics[width=\linewidth]{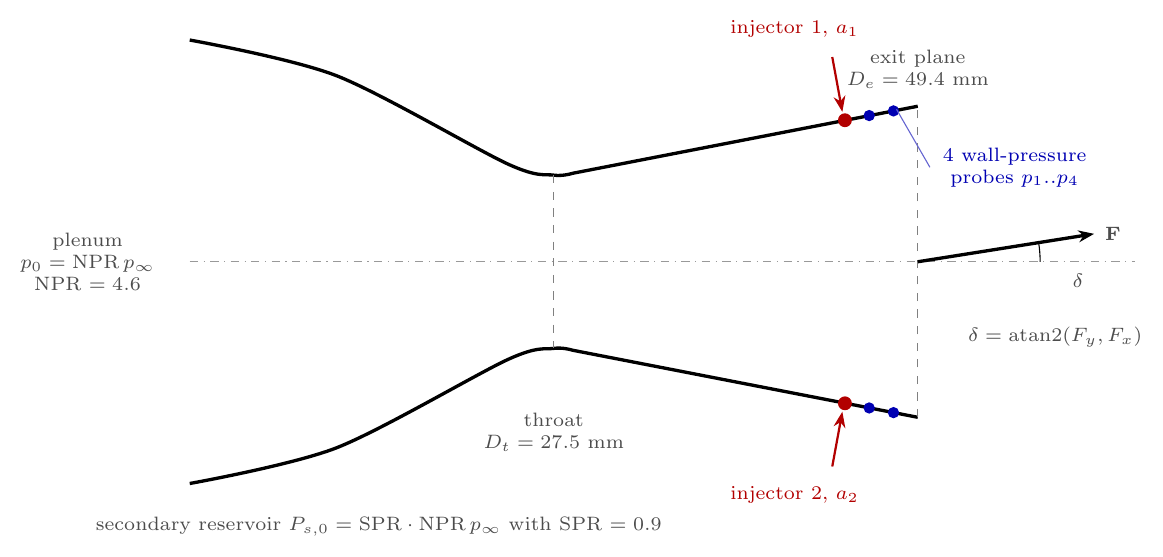}
\caption{Two-dimensional convergent--divergent nozzle of the HydroGym \texttt{Nozzle2D} environment \citep{lagemann2025hydrogym}, drawn to scale from the environment geometry, which follows the NASA Langley experiment of \citet{waithe2003multiple}. One injector sits on each divergent wall; four wall-pressure probes lie between the injectors and the exit plane. The thrust angle $\delta$ is computed from the momentum and pressure flux through the exit plane.}
\label{fig:nozzle}
\end{figure}

The 2D jet engine nozzle control of HydroGym \citep{lagemann2025hydrogym}, a shock-vector-control (SVC) problem, is chosen to try the developed insect-inspired RL network to test how the model performs in 2D flow physics problems. Here, primarily gas flows through a converging-diverging nozzle. Two small injectors in the downstream (one at the top wall and another at the bottom, as in Figure~\ref{fig:nozzle}) introduce secondary gas to the main flow. This additional gas injection will be controlled by the neural networks, which change the pressure and shock pattern, thus allowing the exhaust direction to change. The goal of this benchmark is to control the exhaust thrust vector to a requested angle. 

The environment solves the two-dimensional compressible Euler equations for a calorically perfect gas (i.e., $\gamma_g=1.4$, $R_g=287.14$~J\,kg$^{-1}$K$^{-1}$),
\begin{gather}
\frac{\partial\mathbf{U}}{\partial t}+\nabla\cdot\mathbf{F}(\mathbf{U})=\mathbf{0},\notag\\
\mathbf{U}=\begin{pmatrix}\rho\\ \rho\mathbf{v}\\ \rho E\end{pmatrix},\quad
\mathbf{F}=\begin{pmatrix}\rho\mathbf{v}\\ \rho\mathbf{v}\otimes\mathbf{v}+p\mathbf{I}\\ (\rho E+p)\mathbf{v}\end{pmatrix},\notag\\
p=(\gamma_g-1)\rho\Big(E-\tfrac12|\mathbf{v}|^2\Big),
\end{gather}

using JAX-Fluids \citep{bezgin2023jaxfluids,bezgin2025jaxfluids2}. In this ideal-gas configuration, the viscous effects and the heat fluxes are disabled. It conserves heat, momentum, and energy while resolving downstream shocks. Temperature can still change through compression and expansion, but the policies ignore it. The solver is a Godunov finite-volume scheme that uses the HLLC Riemann solver and WENO5-Z reconstruction of characteristic variables. Time advances with a third-order Runge–Kutta scheme. The nozzle walls are immersed boundaries described by a level set. The environment is a coarse grid with $500\times300$ cells, uniform at 0.55~mm in the nozzle region ($D_t/\Delta x=50$). The plenum is held at $p_0=\mathrm{NPR}\,p_\infty$ with NPR $=4.6$, $p_\infty=100$~kPa and $T_\infty=300$~K.

\subsection{Secondary injection}
Action $a_i\in[0,1]$ sets the total pressure of injector $i$ between the local wall pressure and the secondary reservoir pressure $P_{s,0}=\mathrm{SPR}\cdot\mathrm{NPR}\,p_\infty$. Here the nozzle pressure ratio ($\mathrm{NPR})$ is the primary-jet total pressure divided by the ambient pressure. The secondary pressure ratio ($\mathrm{SPR})=0.9$ is the injector supply pressure divided by the primary-jet total pressure. Therefore, with $p_\infty=100$~kPa, this gives $P_{s,0}=414$~kPa.
\begin{equation}
p_{s,i}=p_{\mathrm{loc}}+a_i\big(P_{s,0}-p_{\mathrm{loc}}\big).
\end{equation}
The injector state follows from isentropic expansion to sonic conditions when $p_{\mathrm{loc}}/p_{s,i}<\big(2/(\gamma_g+1)\big)^{\gamma_g/(\gamma_g-1)}$ (choked), and otherwise to the local pressure, with Mach number
\begin{equation}
M_{\mathrm{inj}}=\sqrt{\frac{2}{\gamma_g-1}\left[\left(\frac{p_{s,i}}{p_{\mathrm{loc}}}\right)^{(\gamma_g-1)/\gamma_g}-1\right]}.
\end{equation}
The resulting mass, momentum, and energy fluxes enter through the nozzle tip of width 2.03~mm that forms each injector.

\subsection{Thrust vector, reward, and observations}
The 2-D thrust vector ($F_x$, $F_y$) is defined by integrating over the exit plane $A_e$,
\begin{gather}
F_x=\int_{A_e}\big[\rho u^2+(p-p_\infty)\big]\,dA,\qquad F_y=\int_{A_e}\rho u v\,dA,\notag\\
\delta=\operatorname{atan2}(F_y,F_x),
\end{gather}
and the reward is the native HydroGym objective $r_t=-|\delta_t-\delta_{\mathrm{ref}}(t)|$. We use the native step target,
\begin{equation}
\delta_{\mathrm{ref}}(t)=\begin{cases}0^\circ, & t\le 0.5~\mathrm{ms},\\ 5^\circ, & t>0.5~\mathrm{ms}.\end{cases}
\end{equation}
The policy acts every 100 solver steps (15~$\mu$s), giving 667 decisions in the 10~ms episode. The observations provided to the policies are current thrust angle ($\delta/\pi$), reference thrust angle ($\delta_{\mathrm{ref}}/\pi$), and two upper and lower wall pressure ($\ p_1/p_0,\dots,p_4/p_0$). The pressure probes sit at one- and two-thirds of the wall segment between each injector and the exit. No observations, reward terms, or episode limits were modified from the source from HydroGym.

\subsection{Insect instance for the nozzle}
The same architecture and layer sizes are used as in the thermal-fluid loop (benchmark \#1, Section~\ref{sec:arch}). The six nozzle readings are passed to the sensory encoder, where they are split into a proprioceptive part (the current and reference thrust angles) and a hazard part (the four wall pressures). The fused code then feeds the mushroom body, central-complex ring, and command center to get its own input, command, and mode probabilities. These are then shared with the 4 local controllers: \texttt{angle\_tracking}, \texttt{transient\_recovery}, \texttt{pressure\_response}, and \texttt{steady\_tracking}. The local action proposals and priorities from the controllers are mixed within the response group and goal group, respectively, by the arbiter. The arbiter then weighs the two groups against each other and maps the result to the two injector commands. The output bias starts at $\mathbf{a}_0=(0.5,0.5)$.

\subsection{Local controllers}
Each controller receives its own input slice plus the shared command and mode probabilities and outputs a 16-D local action proposal and priority score. 
\subsubsection{Angle Tracking}
The local controller `angle\_tracking' receives a 128-D tracking embedding. It reduces the gap between the thrust angle and its target through the injector commands. It belongs to the goal group.
\subsubsection{Transient Recovery}
The local controller `transient\_recovery' receives the fused summary and the ring state, thus becoming the only controller with direct access to the recurrent state. Other controllers receive recurrent information only through the shared command and mode probabilities. It manages the transient effects like overshoot and shock impacts after the sudden change in the expected thrust angle. It belongs to the goal group. 
\subsubsection{Pressure Response}
The local controller 'pressure\_response' receives a 128-D embedding of the four pressure probe readings. It reacts quickly after a pressure overshoot or shock waves. The probes, two on each wall, register the change in the shock pattern rather than the thrust angle. This controller uses that input to keep the environment safe, so it belongs to the response group. 
\subsubsection{Steady Tracking}
The fourth and final local controller receives a narrower 64-D angle embedding and holds the current angle steady once the reference angle settles. It belongs to the response group. 
\section{Training and evaluation}
\label{sec:ppo}

\subsection{Policy optimization}
All policies are trained with proximal policy optimization (PPO)  \citep{schulman2017ppo}. PPO is an actor--critic method where the actor is the policy $\pi_\theta$, and the critic is a value function $V_\theta$ that is used to compute the advantages and to estimate the expected discounted return. In the insect-inspired policy, the critic is a linear head on the command-center GRU state. Similarly, MLP and GRU baselines read their value from the shared trunk through a linear head, while the MoE has a separate critic network (256, 256). Here $\mathbf{s}_t$ denotes the observation along with the policy's recurrent state. With probability ratio $\varrho_t(\theta)=\pi_\theta(\mathbf{a}_t\mid\mathbf{s}_t)/\pi_{\theta_{\mathrm{old}}}(\mathbf{a}_t\mid\mathbf{s}_t)$ and advantage $\hat A_t$, the policy loss is
\begin{equation}
L^{\mathrm{clip}}_t=\begin{cases}
\begin{aligned}[t]&\max\big(-\varrho_t\hat A_t,\\&\quad\,-\mathrm{clip}(\varrho_t,1-\varepsilon,1+\varepsilon)\hat A_t\big),\end{aligned} & \hat A_t\ge0,\\[2pt]
\begin{aligned}[t]&\min\Big(\max\big(-\varrho_t\hat A_t,\\&\quad\,-\mathrm{clip}(\varrho_t,1-\varepsilon,1+\varepsilon)\hat A_t\big),\\&\quad\,-c\hat A_t\Big),\end{aligned} & \hat A_t<0,
\end{cases}
\end{equation}
which is the dual-clip form of \citet{ye2020dualclip} with $c=3$ and $\varepsilon=0.2$. Advantages use generalized advantage estimation \citep{schulman2016gae},
\begin{align}
\delta^V_t&=r_t+\gamma V(\mathbf{s}_{t+1})(1-d_{t+1})-V(\mathbf{s}_t),\notag\\
\hat A_t&=\sum_{l\ge0}(\gamma\lambda)^l\,\delta^V_{t+l},
\end{align}
where $d_{t+1}=1$ if the episode ends at step $t+1$, $\gamma=0.99$ and $\lambda=0.95$. Training rewards are normalized by the standard deviation of the discounted return, which keeps the value targets on a similar scale throughout training. The full loss is
\begin{equation}
L=\overline{L^{\mathrm{clip}}}+0.5\,\overline{(V_\theta-\hat R)^2}-0.001\,\overline{\mathcal{H}[\pi_\theta]}+L_{\mathrm{aux}},
\end{equation}
\begin{align}
L_{\mathrm{aux}}^{\mathrm{Insect}}&=10^{-4}\,\overline{\|\mathbf{c}_t\|_1}-10^{-3}\,\overline{\mathcal{H}(\mathbf{p}_t)}-10^{-3}\,\overline{\mathcal{H}(\mathbf{w}_t)},\notag\\
L_{\mathrm{aux}}^{\mathrm{MoE}}&=-10^{-3}\,\overline{\mathcal{H}(\mathbf{w}^{\mathrm{router}}_t)},
\end{align}
where $\mathcal{H}$ is the entropy and the bar denotes a mini-batch mean. The entropy terms on the mode and module weights act as a weak bonus against early collapse of the routing; they do not reward any particular module. Recurrent policies are trained by backpropagation through 40-step sequences whose initial hidden states are stored during the rollout. Optimization uses Adam \citep{kingma2015adam} ($\epsilon=10^{-5}$) with learning rate $3\times10^{-4}$, four epochs per update, minibatches of 256 transitions and gradient-norm clipping at 0.5.

\subsection{Thermal protocol}
Each architecture is trained from scratch with ten seeds for 400 updates of 16 parallel environments $\times$ 240 steps, which means\ 1.54~million control steps per run, with identical PPO settings for all four architectures. Every 10 updates, the deterministic policy is evaluated on three fixed validation episodes, and the checkpoint with the highest undiscounted validation return is kept. The selected policies are then tested on five held-out groups that were not used for training or selection:
\begin{itemize}
\item \emph{nominal}: 5 new episodes per seed from the training distribution (50 per architecture);
\item \emph{final distribution}: 10 further episodes per seed from the same distribution (100 per architecture);
\item \emph{stress}: heat pulses $\times1.15$, cooling $\times0.90$, branch resistances $\times1.15$, sensor delay of 2 steps (50 per architecture);
\item \emph{forced recirculation}: at $t=250$~s, the risk of one upstream mode, one downstream mode or three modes is set to 1.10, above the unsafe level (90 per architecture);
\item \emph{compositional}: combining five cases—stronger pulses ($\times1.15$), weaker cooling ($\times0.75$), stronger degradation, overlapping vulnerability windows and a sensor delay of 6 steps, one with an added forced-recirculation event (50 per architecture).
\end{itemize}
We evaluate each trained policy deterministically on held-out episodes. The main accuracy metric is the mean-zone RMS temperature error over an episode. We also report the goal-reach rate, the fraction of episodes in which all three zones stay within a tolerance band around the target for 60~s, both at the strict $\pm2$~K band and as the band widens to $\pm3.5$~K; the peak recirculation-risk index; and the energy used over the episode. During training we track the deterministic evaluation return, the pump and valve commands, and the recirculation risk, and, for the Insect policy and the MoE, the entropy of the module weights. The last one is to see how the insect-inspired policy is prioritizing the task and allocating its controller accordingly, rather than equally blending the weights. Learning curves show the median over the ten seeds with the 25--75\% range, and episode metrics are shown for every seed together with the mean across seeds.

\subsection{Nozzle protocol}
Each policy is trained for 40,000 actions per seed (200 updates of 200 actions, about 60 episodes) with five seeds on GPUs at Virginia Tech's Advanced Research Computing Center. The MLP follows HydroGym's Stable-Baselines3 \citep{raffin2021sb3} example, changed only in its width and in the rollout length of 200 actions shared with the Insect policy: PPO defaults (same as above), separate tanh actor and critic networks, a diagonal Gaussian clipped to $[0,1]$, and running observation and reward normalization (clip 10). The insect policy uses the recipe of the thermal study (Section~\ref{sec:ppo}) and receives observations through the same running normalizer. The two policies therefore have matched size and inputs but not identical training recipes. After training, each policy is run deterministically for one full 10~ms episode. Checkpoints saved every 4,000 actions give learning curves. Here, we report the training return, the steady-state thrust-angle error after the step, the overshoot, and the fluctuation of the exit-plane pressure, and compare the two policies seed by seed.

\section{Results}
\suppressfloats[t]

\subsection{Thermal network: learning}

\begin{figure}[!htbp]
\centering
\includegraphics[width=\linewidth]{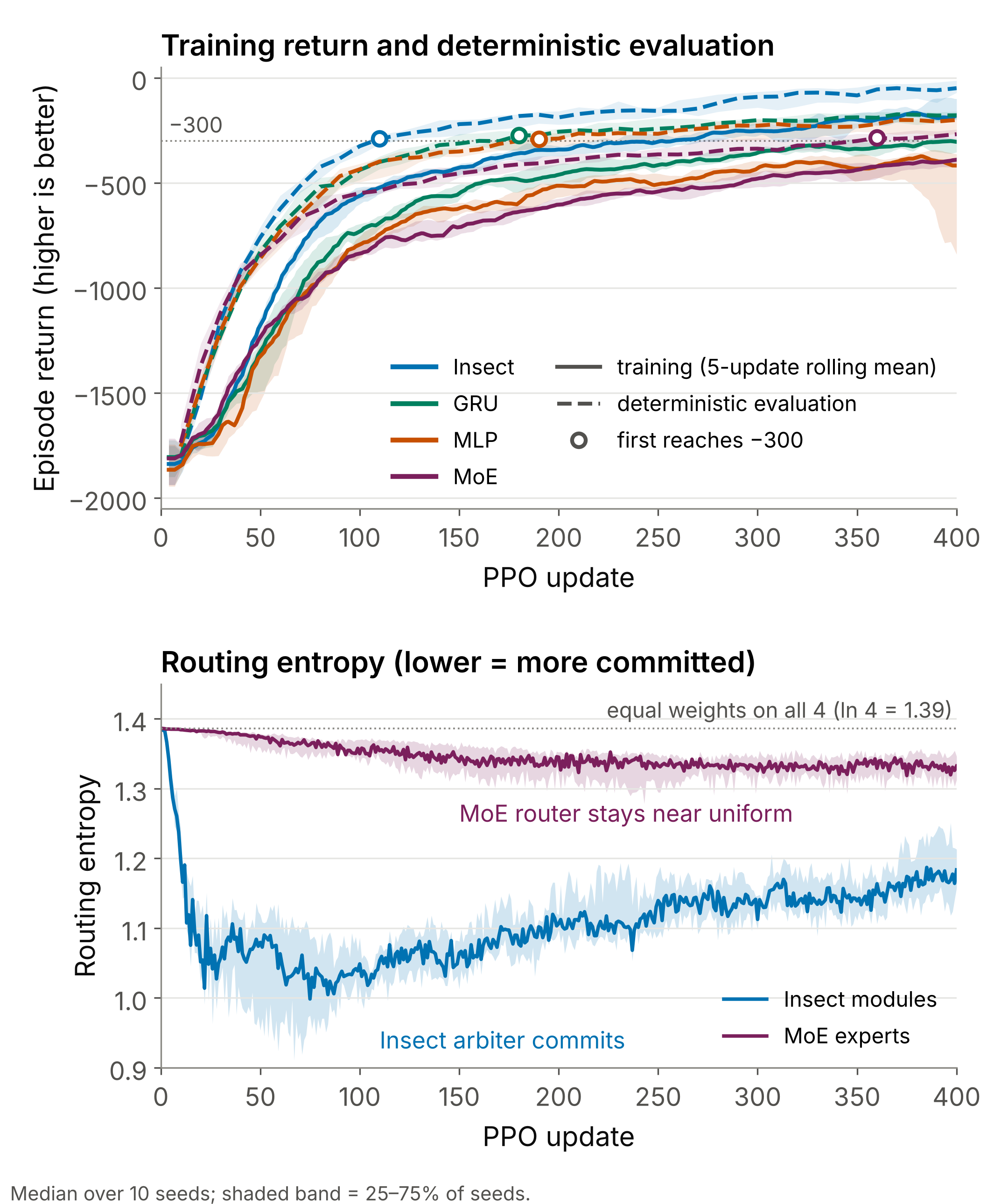}
\caption{Training on the thermal network (median over ten seeds; bands span the 25th to 75th percentiles of the seeds). Top: episode return of the training rollouts (solid, 5-update rolling mean) and of the deterministic evaluation (dashed); circles mark where each return first reaches $-300$. Bottom: entropy of the Insect module weights and of the MoE router weights; equal weights on all four modules give $\ln4\approx1.39$.}
\label{fig:training}
\end{figure}

The insect-inspired policy reached the highest deterministic evaluation return of the four architectures from roughly the 50th update onward (Figure~\ref{fig:training}). Its evaluation return first reached $-300$ at update 110, against updates 180--360 for the baselines. At the final update, the median evaluation return was $-49.7$ for the Insect policy, against $-178.8$ for the GRU, $-200.9$ for the MLP, and $-268.7$ for the MoE. Because the return includes the shaping term, energy usage, and maintaining the zones within the target temperature (the success bonus), it is useful for ranking the policies but has no direct physical meaning. Therefore, we base the comparison on the held-out physical metrics.

The entropy of the Insect module weights fell from $\ln4\approx1.39$ to about 1.0 within the first 75 updates and then rose slowly to about 1.2 by update 400. This proves that the insect-inspired policy is prioritizing the controller actions based on the situation rather than equally blending the weights to all the controllers, like in the MoE router entropy which stayed near 1.33 throughout (Figure~\ref{fig:training}, bottom). Both policies receive the same small entropy bonus on their routing weights (Section~\ref{sec:ppo}), so we can confidently conclude that the Insect arbiter commits to particular modules much more than the MoE router does.

\begin{figure}[!htbp]
\centering
\includegraphics[width=\linewidth]{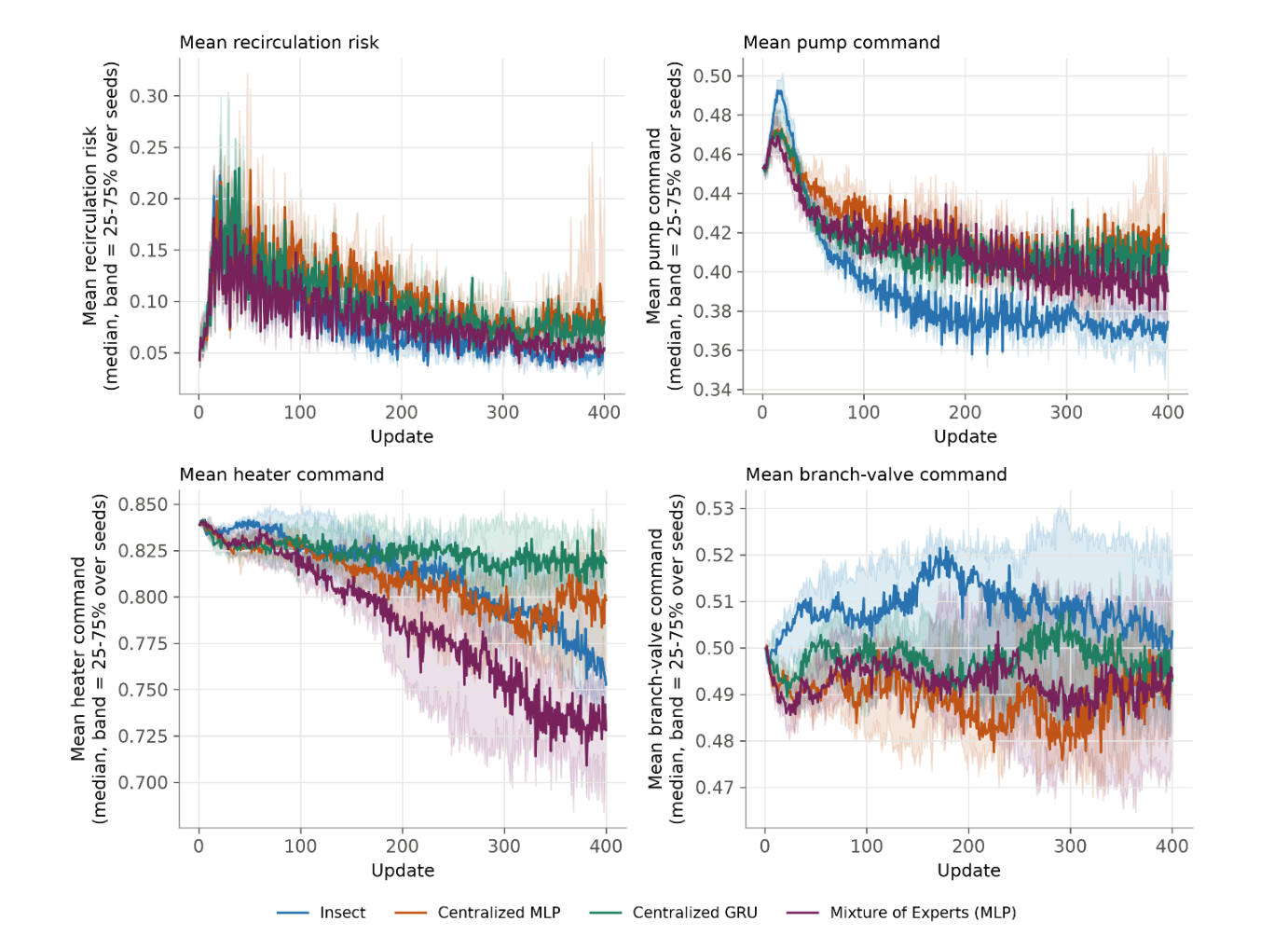}
\caption{Mean recirculation risk and actuator commands in the training rollouts (median over ten seeds; bands span the 25th to 75th percentiles). Commands are scaled to $[0,1]$. Because the valve split depends only on the relative valve openings (Eq.~\ref{eq:split}), the mean valve command (bottom right) does not affect the flow by itself.}
\label{fig:actuators}
\end{figure}

The architectures showed different operating capabilities during training (Figure~\ref{fig:actuators}). At update 400, the median pump command of the Insect policy was 0.37, against 0.40--0.42 for the baselines; with $Q=Q_0+G_pu_p$, this corresponds to 6--9\% less total flow. The Insect policy also kept the lowest mean recirculation risk in its training rollouts (0.048, against 0.061--0.084), and all four architectures stayed far below the unsafe level $R_{\mathrm{unsafe}}=1$.

\subsection{Thermal network: held-out accuracy and safety}

\begin{table*}[t]
\centering
\caption{Nominal held-out results (50 episodes per architecture; mean $\pm$ s.d. over ten seeds).}
\label{tab:nominal}
\small
\begin{tabular}{lcccccc}
\toprule
 & Worst-zone & Mean-zone & Worst-zone & Success & Would-trip & Median time \\
 & RMSE [K] & RMSE [K] & IAE [K\,s] & [\%] & [\%] & to success [s] \\
\midrule
Insect & \textbf{3.37 $\pm$ 0.26} & \textbf{1.44 $\pm$ 0.18} & \textbf{1702 $\pm$ 1041} & \textbf{58} & 0 & 207 \\
MLP & 6.08 $\pm$ 2.87 & 3.15 $\pm$ 0.78 & 5238 $\pm$ 2441 & 2 & 8 & 173 \\
GRU & 4.65 $\pm$ 0.64 & 2.62 $\pm$ 0.45 & 4075 $\pm$ 567 & 0 & 0 & -- \\
MoE & 6.23 $\pm$ 0.58 & 3.86 $\pm$ 0.48 & 5503 $\pm$ 543 & 0 & 0 & -- \\
\bottomrule
\end{tabular}
\end{table*}

\begin{table*}[t]
\centering
\caption{Reduction of worst-zone RMSE achieved by the Insect policy in each held-out group [\%]; mean over ten paired seeds with 95\% confidence interval. Positive values favor the Insect policy.}
\label{tab:relative}
\small
\begin{tabular}{lccc}
\toprule
Held-out group & vs MLP & vs GRU & vs MoE \\
\midrule
Nominal & 38.8 [27.7, 50.0] & 26.4 [19.0, 33.8] & 45.5 [41.2, 49.9] \\
Final distribution & 35.8 [24.3, 47.3] & 21.8 [15.6, 28.0] & 39.0 [34.9, 43.1] \\
Forced recirculation & 31.3 [19.5, 43.0] & 18.0 [8.9, 27.0] & 37.6 [34.1, 41.1] \\
Stress & 34.3 [20.0, 48.5] & 6.2 [$-$0.2, 12.7] & 25.8 [18.6, 32.9] \\
Compositional & 27.7 [11.8, 43.6] & 8.5 [$-$1.0, 18.0] & 12.2 [$-$2.3, 26.7] \\
\bottomrule
\end{tabular}
\end{table*}

On the nominal held-out episodes, the Insect policy had the lowest worst-zone RMSE for every one of the ten seeds against each baseline (as shown in Table~\ref{tab:nominal}). 58\% of the episodes succeeded, against 2\% for the MLP and none for the GRU or MoE. The mean reduction was 38.8\% relative to the MLP, 26.4\% relative to the GRU, and 45.5\% relative to the MoE (as shown in Table~\ref{tab:relative}). The in-between run spread was also smallest for the insect-inspired policy (0.26~K), while one MLP seed reached 14.0~K and accumulated 359~s above the unsafe recirculation level.

\begin{figure}[!htbp]
\centering
\includegraphics[width=\linewidth]{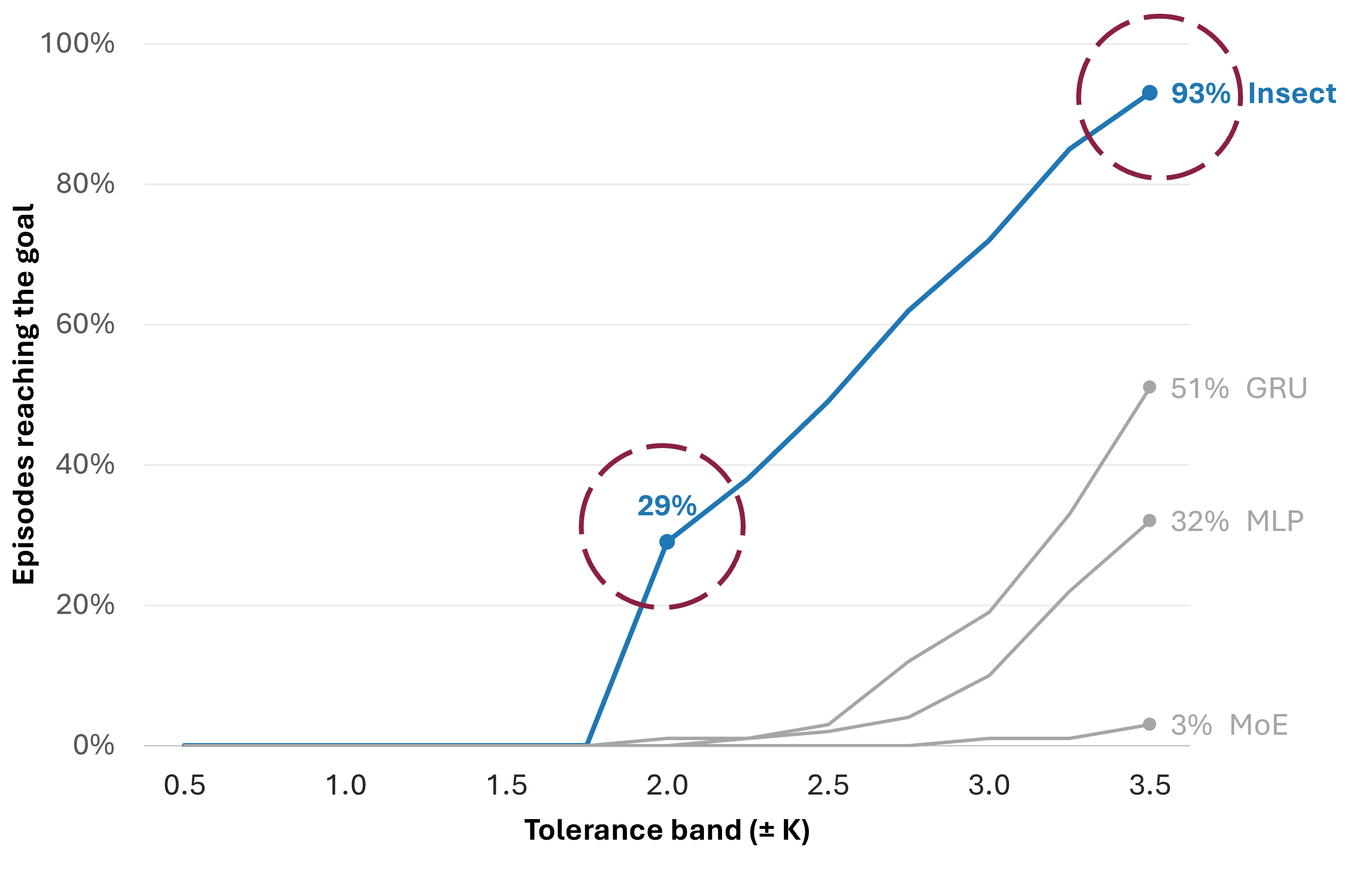}
\caption{Share of final-distribution held-out episodes (100 per architecture) that reach the goal when the success rule is replayed with a wider or narrower tolerance band. Episodes end once the trained $\pm2$~K goal is met, so the values for tighter bands (shaded) are lower bounds.}
\label{fig:goalreach}
\end{figure}

The Insect policy was the only architecture that regularly reached and held the goal of all three zones within $\pm2$~K for 60~s. This test was designed to check for the network's ability to maintain the plant in the required run conditions. In the final-distribution group, the rates were 29\% for the Insect policy, 1\% for the MLP, and 0\% for the GRU and MoE (Figure~\ref{fig:goalreach}). Then, expanding the goal band from $\pm2$~K to $\pm3.5$~K to see at what point the baseline can maintain and win the objective. 93\% of the Insect episodes reached the goal, against 51\% for the GRU, 32\% for the MLP, and 3\% for the MoE. Even at the broader goal band, the insect-inspired network outperformed the baselines.

\begin{figure}[!htbp]
\centering
\includegraphics[width=\linewidth]{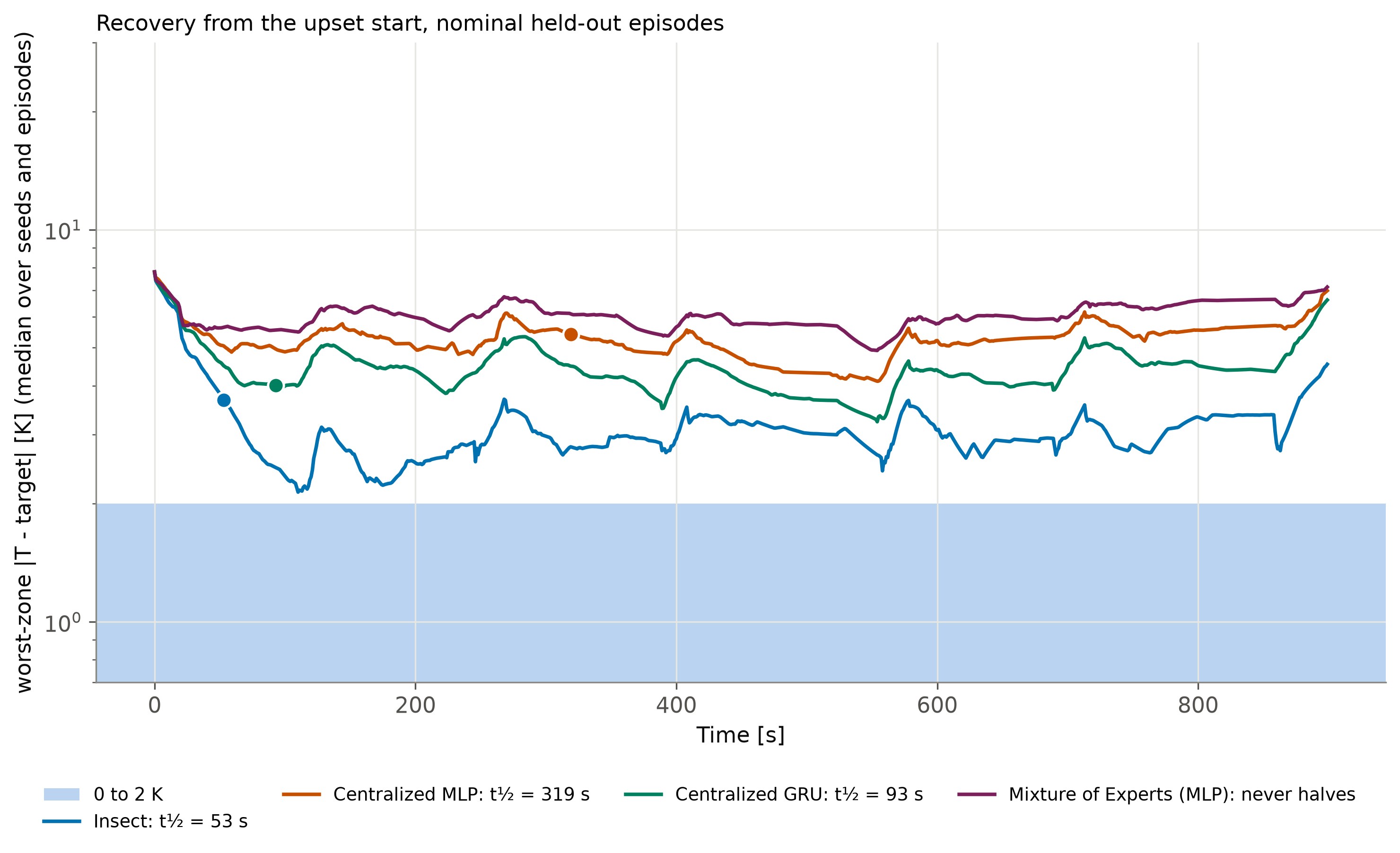}
\caption{Worst-zone error after the upset start in the nominal held-out group (median over seeds and episodes, log scale). Markers give the time at which each median error first halves.}
\label{fig:recovery}
\end{figure}

The Insect policy also recovered fastest from the upset start (Figure~\ref{fig:recovery}). With every episode initially starting at 4--9~K off the zone's target temperature, the networks have to bring the error closer to 0. In the nominal group, the insect-inspired model's median worst-zone error halved in 53~s, against 93~s for the GRU and 319~s for the MLP, while the MoE did not halve its error within the 900~s episode.

\begin{figure*}[t]
\centering
\includegraphics[width=\textwidth]{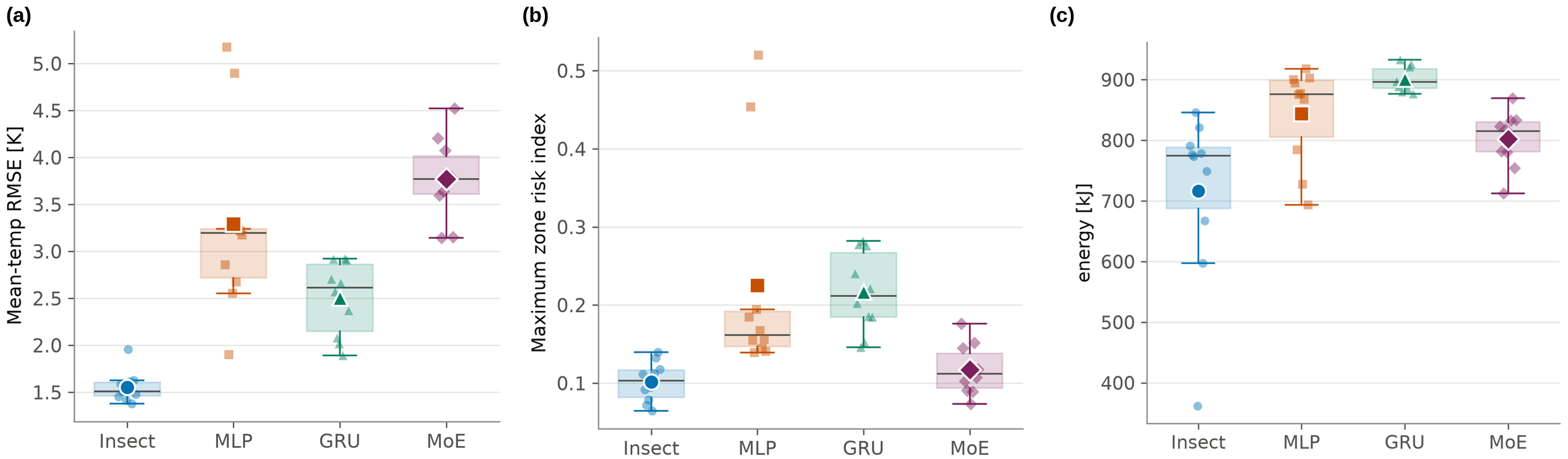}
\caption{Final-distribution held-out group, one small marker per training seed (mean over its ten episodes): (a) RMS error of the mean zone temperature, (b) peak recirculation risk over the episode, (c) actuator energy per episode. Boxes span the 25th to 75th percentiles of the seeds, horizontal lines give the median, large markers the mean, and whiskers extend to 1.5 times the interquartile range.}
\label{fig:boxplots}
\end{figure*}

Figure~\ref{fig:boxplots} summarizes the final-distribution group per training seed. The median mean-zone RMSE of the Insect policy was 1.5~K, against 2.6--3.8~K for the baselines, which is \ 40--60\% lower. At the same time, the median peak recirculation risk was also the lowest for the insect-inspired network (0.10, against 0.11--0.21), although the MoE came close (0.11). The median energy consumed by the network to run the plant per episode was 5--14\% lower than that of the baselines (815--897~kJ).

\subsection{Thermal network: robustness}
The advantage narrowed as the test conditions moved away from the training distribution (Table~\ref{tab:relative}). Under stress, the Insect policy still improved on the MLP and MoE, but its gain over the GRU (6.2\%) was not distinguishable from zero. 42\% of the insect's stress episodes reached a trip condition, the same rate as the GRU (while MLP 80\%, and MoE 66\%). In the compositional group every architecture tripped in every episode, and only the improvement over the MLP was clearly positive. These cases lie outside the training distribution, and none of the policies handle them.

The time for the injected risk to fall halfway back to its pre-event level was $21.2\pm2.7$~s for the Insect policy, against $16.3\pm2.6$~s (MLP), $17.9\pm3.3$~s (GRU) and $13.8\pm1.8$~s (MoE). Its benefit in these episodes came from regulating temperature during and after the event, not from faster suppression.

\subsection{Thermal network: arbitration}

\begin{figure*}[t]
\centering
\includegraphics[width=0.9\textwidth]{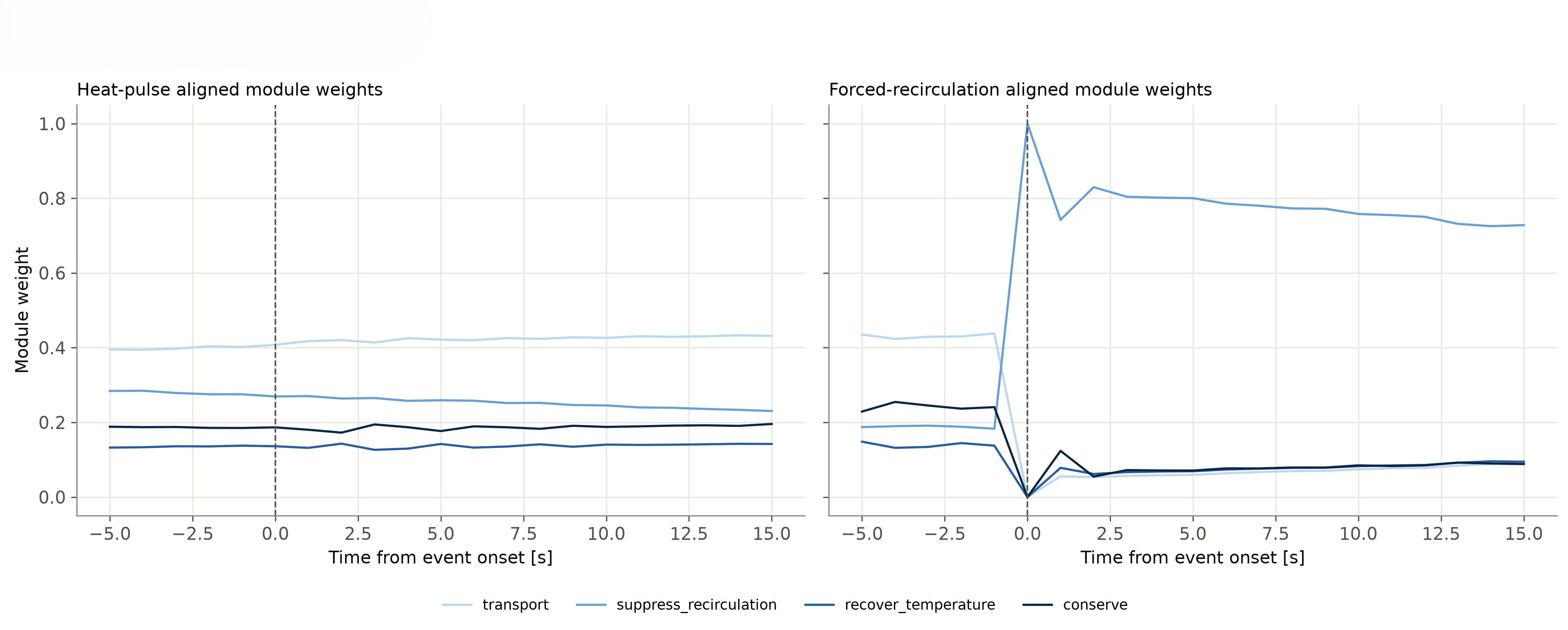}
\caption{Module weights $w_j$ of a final deterministic Insect policy aligned to the onset of heat pulses (left) and of a forced-recirculation event (right), highlighting how weights are prioritized based on the situation and requirement.}
\label{fig:modules}
\end{figure*}

Figure~\ref{fig:modules} shows how a final Insect policy uses its modules. In a normal operation, the transport module carries about 0.4 of the weight, and the other three share the rest from 1.0. When recirculation is intentionally forced, the weight moves almost entirely to the suppress-recirculation module at onset (close to 1.0, then about 0.8 for the next 15~s), the module-weight entropy drops from about 1.3 to near zero, and the mixture returns to its normal state over roughly 50~s as the risk decays. The policy cannot observe heat pulses directly, so any response to them has to come through their effect on the measured temperatures. In here it leaves the mixture nearly unchanged. This might be due to insignificant effects of the heat pulse itself on the plant performance. Thus, this underscores a key feature of the insect algorithm that it prioritizes the controller action that will help in reducing the degradation of plant performance. It learned the factors needed for a safer plant and accordingly distribute the weights. The weights $w_j$ are shared by all actuators, so the switch changes the commands to the pump, heater, and valves together.

\subsection{Thermal network: ablations}

To find out which parts of the Insect policy matter, we trained seven ablated variants with ten seeds each and the same PPO settings and compared them with the full un-ablated policy. A1--A7 define the respective ablations and changes made to the original insect-inspired architecture:
\begin{itemize}
\item A1 completely removes the mushroom-body component, including its read-out and Kenyon-cell activity, and sets traces to zero. Therefore, no execution of mushroom-body projection, top-K operation, trace updates, readouts, or valence pathway. 
\item A2 completely removes the central-complex ring state from both command-center passes and from the local controllers. In this run, we will supply zero central-complex state to the first command center pass, and any incoming heading state are ignored.
\item A3 removes adaptive specialists arbitration by fixing all four module weights at $1/4$, bypassing the priorities and the two-level arbiter. Here, we disconnected the learned priorities and the hierarchical group gate. Their parameters are registered to preserve initialization and nominal parameter counts, but they will have no influence on the mixture and receive no gradients. We will retain \texttt{module\_sparse\_coef} and entropy logging as it was.  
\item A4 shuffles the existing input features of the specialists among themselves, so each specialist controller gets a different mixture, but still each controller receives a fixed but scrambled input. For instance, the original architecture has a pre--defined input flow to their respective controllers, but in this ablation study that will be randomized. So we will collect the features generally given to the four specialists, rearrange them using a fixed mapping, and distribute them back to the specialists. The same feature assignment will be applied at every timestep, across all 10 training seeds, and during evaluation. 
\item A5 replaces the four local controllers with one shared controller that produces all four proposals and priorities. It is widened (to 244) to keep the total parameter count approximately unchanged.
\item A6 resets the command-center, central complex ring, and mushroom-body memory states at every decision, so no learned state is carried from one decision to the next. The two command-center passes within a decision remain, and the observation still contains the recent commands and the time held in the band. This is done by starting the mushroom-body Kc trace from 0 at the beginning of every decision. 
\item A7 keeps only the four controller roles, carried out by four MLPs, each reading the raw observation channels behind the corresponding Insect controller, combined by an MoE-style router, with no encoder, memory, ring state, command center, or two-level arbiter.
\end{itemize}
In A1--A3, the removed parts stay in the network but receive no input and no gradient, so these variants have fewer active parameters than the full policy.

\begin{table*}[t]
\centering
\caption{Ablations of the Insect policy in the thermal network and the three baselines, as changes relative to the full Insect policy (mean over ten seeds). IAE [\%] and energy [\%]: worst-zone IAE and actuator energy per episode in the final-distribution group (for actual insect results see Table~\ref{tab:nominal}). Measured recovery: share of the initial worst-zone error removed within 120~s (median over the nominal episodes of each seed; for actual insect results, see Figure~\ref{fig:recovery}). Early and final return: deterministic evaluation return averaged over updates 10--200 and 360--400. For reference, the full policy scores 2839~K\,s, 716~kJ, 69\%, $-539$ and $-52$ on these five measures. Overall, (+) means the values are deviating on the bad side in case of IAE and energy, and (-) in the other three columns means the ablations are doing worse. Bold: $p<0.05$ in a two-sided Wilcoxon signed-rank test, with seeds paired with the full policy.}
\label{tab:ablation}
\small
\setlength{\tabcolsep}{5pt}
\begin{tabular}{llrrrrr}
\toprule
 & & \multicolumn{5}{c}{Change relative to the full Insect policy} \\
\cmidrule(l){3-7}
 & Variant & IAE [\%] & Energy [\%] & Recovery [pp] & Early return & Final return \\
\midrule
\multicolumn{7}{l}{\textit{One component removed or changed}} \\
A1 & No mushroom-body memory & $+5$ & $+6$ & $-2$ & $-6$ & $-39$ \\
A2 & No central-complex ring state & $+3$ & $+5$ & $\mathbf{-8}$ & $-1$ & $-20$ \\
A3 & Uniform arbitration & $+0.5$ & $+3$ & $0$ & $\mathbf{-59}$ & $-8$ \\
A4 & Scrambled controller inputs & $+11$ & $+14$ & $\mathbf{-8}$ & $\mathbf{-41}$ & $-44$ \\
A5 & One shared controller & $+19$ & $+14$ & $-10$ & $\mathbf{-40}$ & $-66$ \\
A6 & No temporal state & $+12$ & $\mathbf{+12}$ & $-5$ & $\mathbf{-31}$ & $-30$ \\
\midrule
\multicolumn{7}{l}{\textit{All Insect components removed, four roles kept}} \\
A7 & Roles only (role MLPs) & $\mathbf{+86}$ & $+13$ & $\mathbf{-39}$ & $\mathbf{-120}$ & $\mathbf{-196}$ \\
\midrule
\multicolumn{7}{l}{\textit{Baselines}} \\
 & MLP & $\mathbf{+97}$ & $\mathbf{+18}$ & $\mathbf{-45}$ & $\mathbf{-110}$ & $\mathbf{-229}$ \\
 & GRU & $\mathbf{+53}$ & $\mathbf{+26}$ & $\mathbf{-29}$ & $\mathbf{-71}$ & $\mathbf{-131}$ \\
 & MoE & $\mathbf{+99}$ & $+12$ & $\mathbf{-48}$ & $\mathbf{-146}$ & $\mathbf{-235}$ \\
\bottomrule
\end{tabular}
\end{table*}

Every single-component ablation increased the worst-zone IAE and the energy per episode in the final-distribution group, but by very small amounts (Table~\ref{tab:ablation}). Among them, one shared controller (A5) cost the most ($+19\%$ IAE), followed by removing the temporal state (A6, $+12\%$) and scrambling the controller inputs (A4, $+11\%$). Removing the mushroom body or the ring state raised the IAE by only 5\% and 3\%, and uniform arbitration by 0.5\%. None of A1--A6 changed the return at the end of training significantly. The single components mattered mainly for how quickly the policy learned and recovered, as seen in the uniform arbitration (A3) lowering the return over the first half of training by 59, the largest drop of any single ablation, and A4--A6 by 31--41, while A2 and A4 removed 8 percentage points [pp] less of the initial error within 120~s. On the other hand, when multiple components are removed at once, (A7) failed substantially across the board. Its IAE was 86\% higher than that of the full policy, and its failures in initial recovery, learning speed, and final return ($-39$, $-120$ and $-196$). Mainly, multiple-component ablation pulled the insect-inspired architecture to fall within the ranges of the MLP, GRU, and MoE baselines.

\subsection{Nozzle: learning and step tracking}

\begin{figure}[!htbp]
\centering
\includegraphics[width=\linewidth]{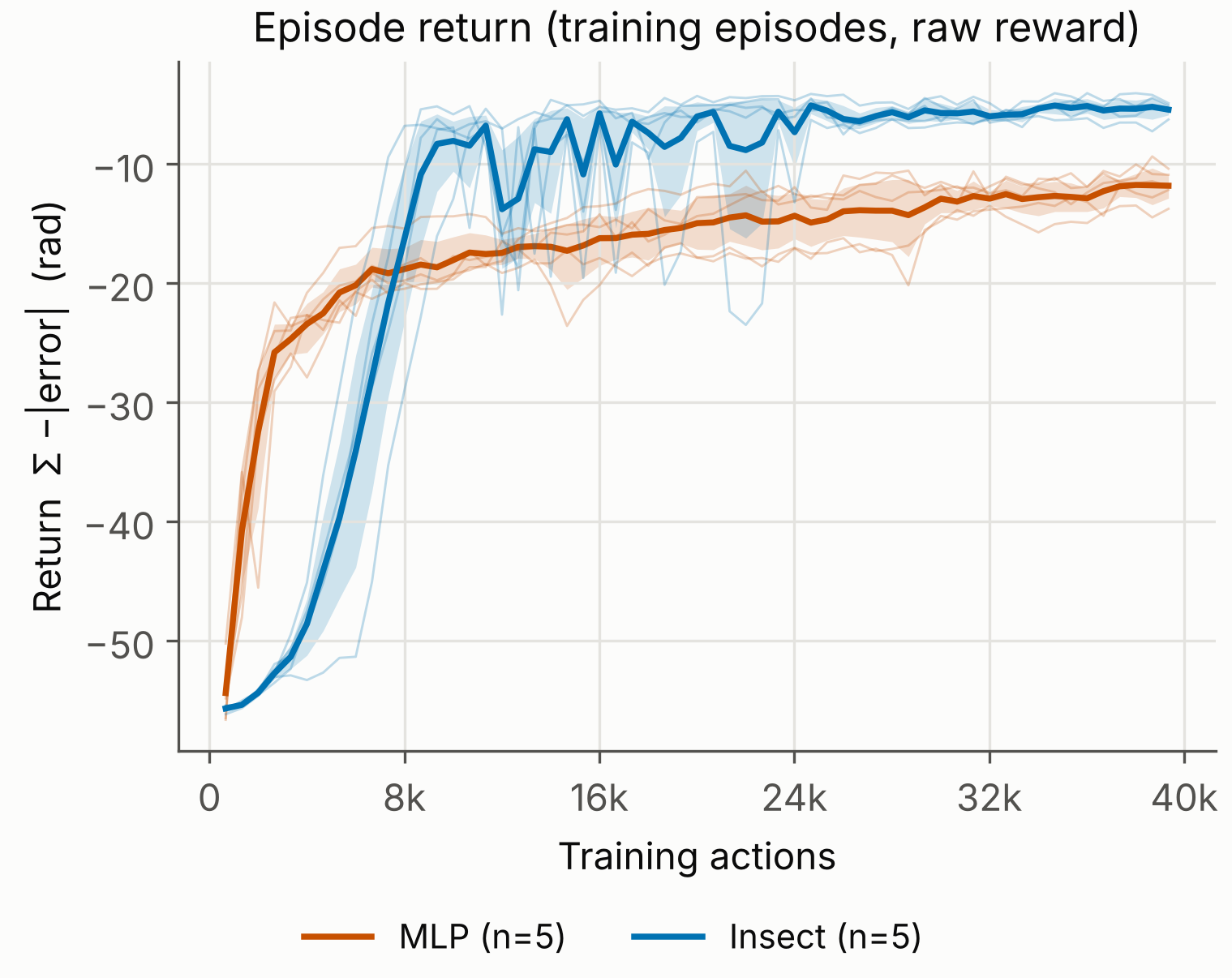}
\caption{Episode return on the nozzle during training, $\sum_t-|\delta_t-\delta_{\mathrm{ref}}|$ in radians over the 667 decisions of an episode (mean and 95\% confidence interval over five seeds; thin lines are single seeds).}
\label{fig:nozzle_return}
\end{figure}

The MLP improved faster over the first few thousand training actions, but the Insect policy overtook it near 8,000 actions (as in Figure~\ref{fig:nozzle_return}). At the end (40,000 actions), the training return was about $-5$ for the Insect policy and $-12$ for the MLP. This can also be interpreted as a mean angle error of about $0.4^\circ$ and $1.0^\circ$ per decision for Insect and MLP, respectively.

\begin{table*}[t]
\centering
\caption{Nozzle step-tracking results for the final deterministic policies (mean $\pm$ s.d. over five seeds). The target angle steps from $0^\circ$ to $5^\circ$ at 0.5~ms. The fractions within the parentheses of settled or reached $1^\circ$ are the number of seeds succeeded out of the total seeds; the values are averaged over those seeds.}
\label{tab:nozzle}
\small
\begin{tabular}{lcc}
\toprule
 & Insect & MLP (HydroGym PPO) \\
\midrule
Mean angle error, whole episode [deg] & \textbf{0.39 $\pm$ 0.16} & 1.01 $\pm$ 0.18 \\
Angle error, last 2~ms [deg] & \textbf{0.22 $\pm$ 0.11} & 0.94 $\pm$ 0.22 \\
Overshoot past the $5^\circ$ target [\%] & \textbf{14.4 $\pm$ 7.9} & 56.0 $\pm$ 3.4 \\
Time to settle within $\pm0.5^\circ$ [ms] & 5.9 $\pm$ 4.2 (5/5) & 9.5 (3/5) \\
Training actions to reach $1^\circ$ error & \textbf{9\,600 $\pm$ 2\,190} (5/5) & 27\,800 $\pm$ 7\,300 (4/5) \\
Mean angle error during training [deg] & \textbf{1.04 $\pm$ 0.13} & 1.52 $\pm$ 0.14 \\
\bottomrule
\end{tabular}
\end{table*}

\begin{figure*}[tbp]
\centering
\includegraphics[width=0.9\textwidth]{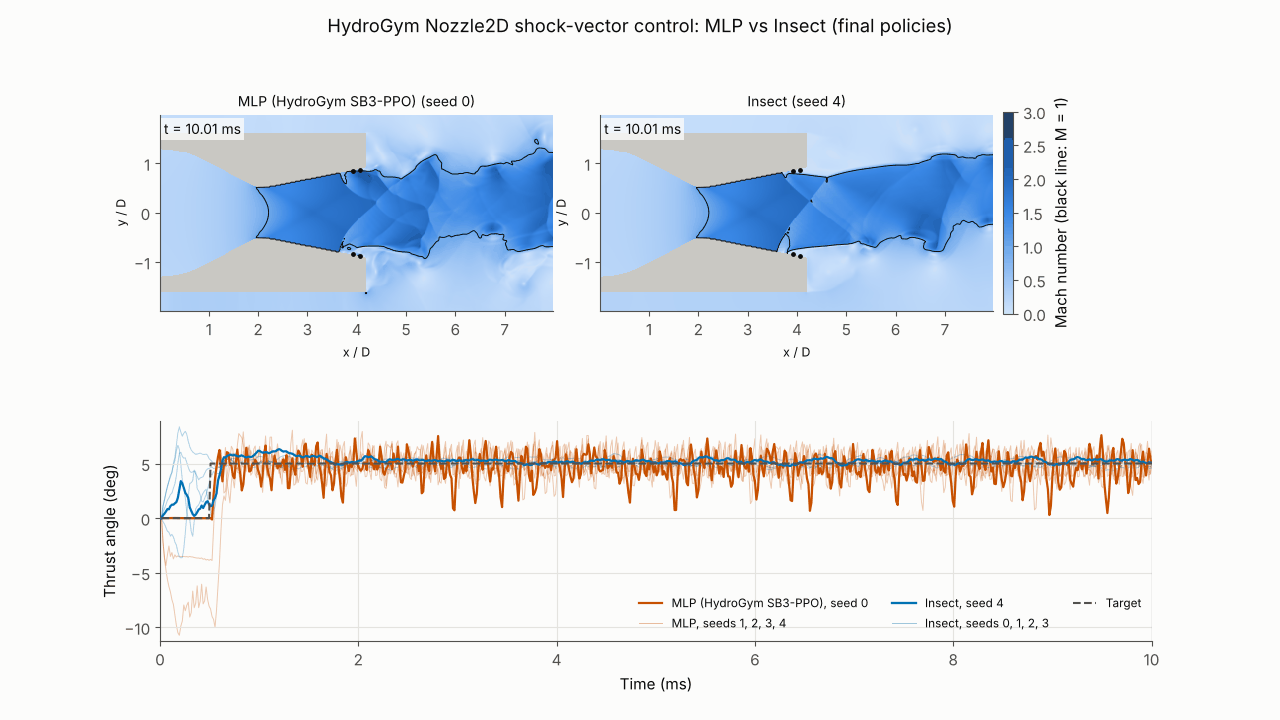}
\caption{Final deterministic policies on the nozzle. Top: Mach number at the end of the episode ($t=10.01$~ms) for the median MLP seed (seed 0) and the median Insect seed (seed 4); the black line marks $M=1$. Bottom: thrust angle of all seeds (thin lines) with the same two seeds highlighted; the target steps from $0^\circ$ to $5^\circ$ at 0.5~ms.}
\label{fig:nozzle_flow}
\end{figure*}

The final deterministic Insect policy tracked the $5^\circ$ step with a mean error of $0.39\pm0.16^\circ$, against $1.01\pm0.18^\circ$ for the HydroGym's reference MLP, and every Insect seed was more accurate than every MLP seed (Table~\ref{tab:nozzle}). The gap was even larger in the final steady state ($0.22^\circ$ against $0.94^\circ$ over the last 2~ms) and in overshoot (14\% against 56\%). The Insect policy also reached $1^\circ$ accuracy sooner in all its five seeds around $9,600\pm2\,190$ training actions, whereas four of five MLP seeds needed $27,800\pm7\,300$, with one MLP seed not reaching it within 40,000. The MLP held the mean angle near the target but oscillated persistently for several degrees, while the Insect policy maintained it close to the target after the sudden step change from $0^\circ$ to $5^\circ$ (Figure~\ref{fig:nozzle_flow}). Once the target reaches $5^\circ$, it stays that way throughout; therefore it is easy for the policy with memory to recall the previous commands and begin from there. While MLP on the other hand, being centralized without memory, couldn't anticipate the target angle and previous command, thus letting its injector commands start from 0. Figure~\ref{fig:nozzle_flow} also includes the flow simulation snapshot at the end of the episode (on top). It depicts the side-by-side comparison of how well the supersonic exhaust flow is maintained by the baseline MLP and our insect-inspired policy. It can be observed that the flow pattern is little unstable when the MLP is controlling the exhaust thrust compared to when it is controlled by the insect-inspired policy. This fluctuation can further be seen in the exit-plane pressure profile.  

\begin{figure}[!htbp]
\centering
\includegraphics[width=\linewidth]{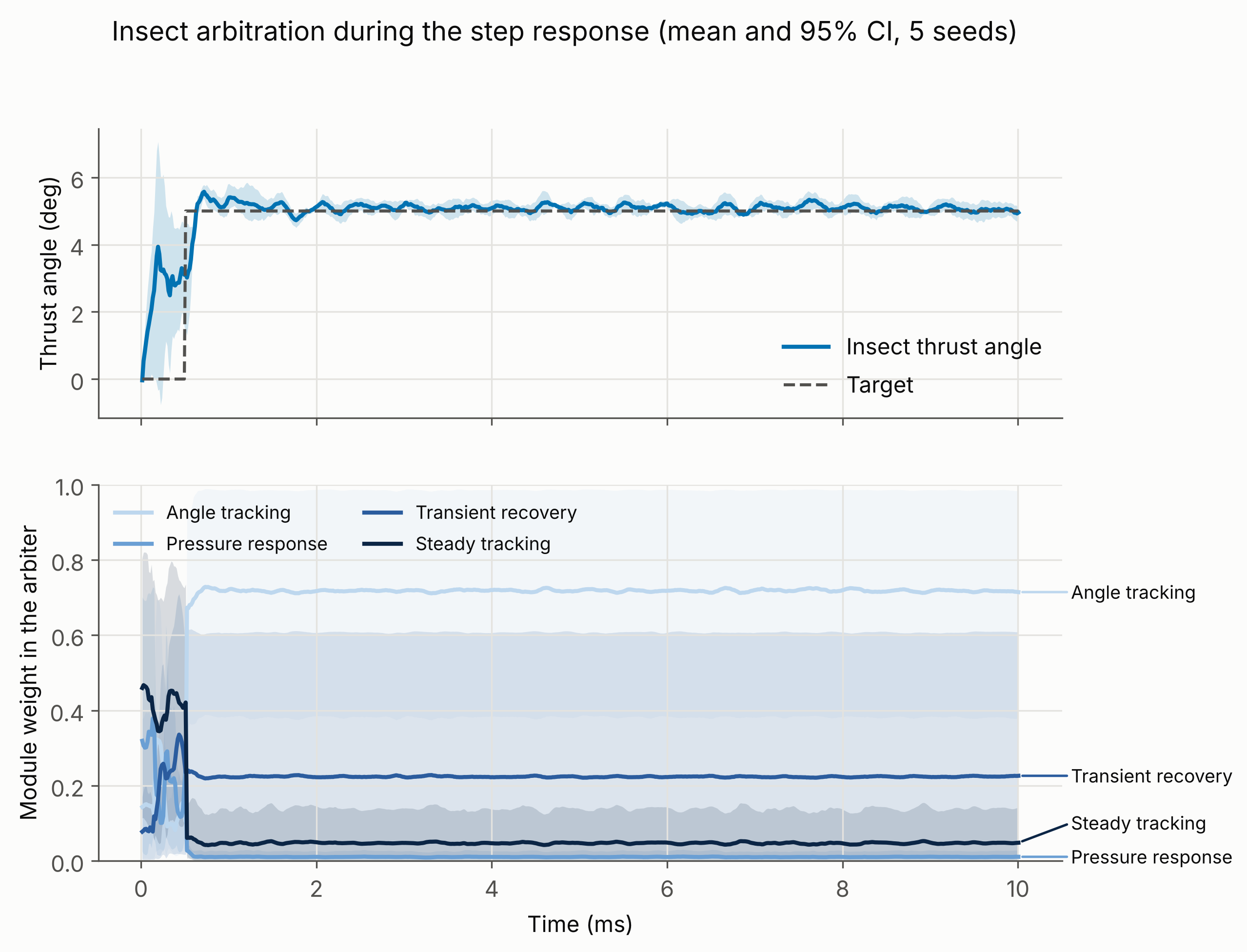}
\caption{Insect policy during the nozzle step response (mean and 95\% confidence interval over five seeds). Top: thrust angle and target. Bottom: arbiter weight of each module. The mean curves blend different module choices across seeds.}
\label{fig:nozzle_arbitration}
\end{figure}

Figure~\ref{fig:nozzle_arbitration} shows how the Insect policy distributes its module weights on the nozzle. During the first 0.5~ms, before the step, the weights move between modules. At the step, one module took about 95--100\% of the weight in four of the five seeds and kept it for the rest of the episode. Therefore, showing the similar prioritizing feature of the insect-inspired policy that emphasizes optimal performance. Here, at the bottom of Figure~\ref{fig:nozzle_arbitration}, angle tracking in three seeds and transient recovery in one of the five seeds are prioritized based on the action requirement within the environment. When there is a competing control command, for instance, a higher bottom injection command will push the exhaust to +5; however, it might also induce pressure outbursts. Therefore, any policy in command should learn the best way to control this. So far, the insect-inspired policy has shown significant results that advocate this control capacity. The mean curves are about 0.72 for angle tracking and 0.22 for transient recovery after the step; therefore, they are averaged over different choices. 

\begin{figure*}[t]
\centering
\includegraphics[width=0.85\textwidth]{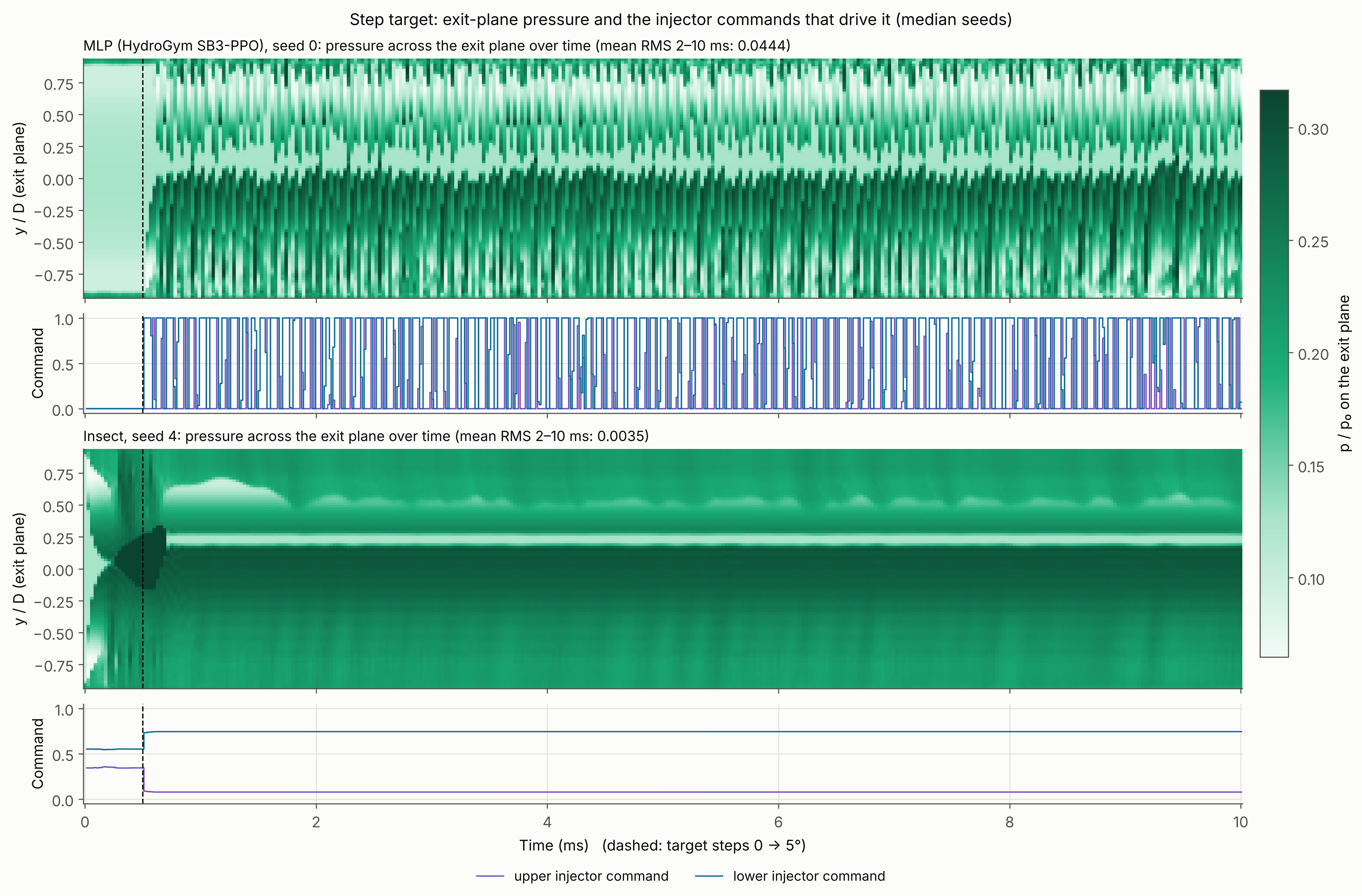}
\caption{Pressure $p/p_0$ across the exit plane over time and the injector commands that produce it, for the median MLP seed (top) and the median Insect seed (bottom). The dashed line marks the target step at 0.5~ms.}
\label{fig:nozzle_exit}
\end{figure*}

The two policies held the target in different ways (Figure~\ref{fig:nozzle_exit}). The MLP switched both injectors fully on and off throughout the episode, and the exit-plane pressure fluctuated wildly. The insect policy settled after the step on nearly constant commands (about 0.75 for the lower injector and 0.08 for the upper one in seed 4), and the exit-plane pressure became almost steady. Between 2 and 10~ms, the RMS fluctuation of $p/p_0$, averaged over the exit plane, was about 0.0035 for Insect seed 4 against 0.0444 for the MLP seed 0. Over the five seeds, the median fluctuation was about eight times lower for the Insect policy, and every Insect seed was below every MLP seed on average.

\section{Discussion}

Across both problems, the Insect policy was more accurate than parameter-matched centralized and modular baselines. In the thermal network it was the only architecture that reliably reached and held the goal, indicating that the insect-inspired policy would be a great supervisory model for optimally cooling down systems. It evidently showed that the network could handle unforeseen disturbances and learn to use minimal energy to keep the plant in a safer zone. The GRU, which has memory but no modules, was the strongest baseline there, and the MoE, which has learned routing but no memory, was the weakest in benchmark \#1. The MoE router stayed close to uniform throughout the training; however, the Insect arbiter committed early to particular modules in the thermal network, displaying higher cognition in prioritizing tasks based on the needs. Similarly, in benchmark \#2, four of five nozzle seeds have prioritized the weights to a single module at the step.

So, the insect-inspired policy could single-handedly beat the centralized baseline with memory (i.e., GRU) and the baseline with modular specialists (i.e., MoE). The arbiter also concentrated the weights on the module that the plant state called for (prioritizing) instead of spreading it evenly (averaging as seen commonly in modular architectures), similar to a typical dust-covered fruit fly, whose cleaning programs are all triggered at once but start with the head, then go along with the cleaning process through a suppression hierarchy \citep{seeds2014grooming}. 
The ablations help to explain where this advantage comes from. No single component was indispensable. Removing any one of them raised the worst-zone IAE by 0.5--19\%, which is 
 not significantly bad when compared to the baselines. And it did not change the final return as much, which suggests that the components partly compensate for one another. The highest single costs came from merging the four local controllers into one (A5), removing the learned state carried between decisions (A6) and scrambling the controller inputs (A4). Judged by the IAE, separate controllers with dedicated inputs matter more for accuracy than either the mushroom-body memory or the ring state on its own. The learned arbitration mattered mainly for learning speed. This was explicitly seen in uniform weights (A3), where the final error was almost unchanged, but the return over the first half of training fell more than for any other single ablation. 
 So far, the single-component ablation did not provide a solid explanation of which feature helps the insect-inspired policy more. Therefore, a combined component ablation was carried out by removing the encoder, memory, ring state, command center, and learned arbitration together (A7), bringing the policy back to the level of the standard networks, so the four controller roles alone do not explain the gain.

On the nozzle, the policies differed as much in how they held the target as in the size of the error. The MLP met the target on average by switching its injectors fully on and off, which kept the exit flow unsteady, whereas the Insect policy settled on a steady injection split. The two policies were trained with different recipes and action distributions to maintain the integrity and authenticity of the algorithm borrowed directly from HydroGym. This limits us from separating the effect of the architecture from that of the training recipe. The pressure-response module, which reads only the wall-probe pressures, received almost no weight in any seed, so the policy steered mainly through the modules driven by the angle readings. This is understandable, since the benchmark defined its rewards only with the angle reading error, not on the pressure deviations. 

Several limitations that need to be addressed in future studies:
\begin{itemize}
\item The thermal plant is a reduced-order model with normalized hydraulic quantities. Though the parameters were fair, they were chosen to create the simple dynamics, not calibrated against a facility.
\item The advantage of the Insect policy was largest within the training distribution. It narrowed under stress, where the Insect policy still outperformed the MLP and MoE but matched the GRU, and in the compositional group, where every architecture failed. Training on harsher and combined disturbances is therefore the most direct way to extend it.
\item On the nozzle, the Insect policy and the reference MLP were each trained with their own recipe, on an inviscid flow with five seeds. Training both with a common recipe, adding ablations, and moving to viscous and three-dimensional simulations will separate the contribution of the architecture more cleanly.
\end{itemize}

Taken together, the two benchmarks show that one distributed architecture, with the same structure and layer sizes, improved control on problems whose control intervals differ by nearly five orders of magnitude. This makes it a promising starting point for flow control under hidden, competing hazards.
\section{Conclusion}

We introduced an RL policy that combines a sparse, multi-timescale Kenyon-cell memory, a ring-structured recurrent state, a recurrent command center, and arbitrated local controllers, following the distributed organization of insect neural control. This fruit fly-inspired neural architecture in a thermally coupled recirculation network with hidden disturbances reduced the worst-zone RMS temperature error by 26--46\% relative to parameter-matched MLP, GRU, and mixture-of-experts baselines in the nominal held-out group and the median mean-zone error by 40--60\% in the final-distribution group. It reached and held the $\pm2$~K goal in 58\% and 29\% of these episodes, where no baseline exceeded 2\%. Importantly, it kept the lowest recirculation risk with about 6--9\% less pump flow during training compared to other baselines. Ablations showed that no single component is a reason for the insect-inspired policy's success; instead, collectively removing insect-inspired features together pulls down the performance and returns of the policy to the level of the standard networks. On HydroGym's supersonic shock-vector-control nozzle, the same architecture tracked a $5^\circ$ thrust-angle step with a mean error of $0.39^\circ$ against $1.01^\circ$ for the reference PPO policy. It also reached $1^\circ$ accuracy with about a third of the training actions, and held the angle with steady injection. Thus, keeping an exit-plane pressure about eight times steadier compared to HydroGym's baseline MLP. Next steps are (i) nozzle ablations with the same training recipe for all policies, (ii) viscous and three-dimensional nozzle simulations, and (iii) a spatially resolved thermal loop and a bench-scale liquid-cooling rig to test the controller on hardware.

\bibliographystyle{unsrtnat}
\bibliography{refs}

\end{document}